\documentclass[prl,twocolumn,superscriptaddress]{revtex4-1}
\usepackage{}
\usepackage{amssymb}
\usepackage{amsfonts}
\usepackage{graphics,graphicx,epsfig,bm,amsmath,amsthm,amssymb}
\usepackage{bm}
\usepackage{bbm}
\usepackage{longtable}
\usepackage{multirow}
\usepackage{array}
\usepackage{color}
\usepackage[usenames,dvipsnames]{xcolor}

\usepackage{float}
\usepackage[a4paper,colorlinks=true,
linkcolor=blue,citecolor=blue,
pdfauthor={ },
pdftitle={ },
pdfsubject={ },
pdfkeywords={ }]{hyperref}

\begin{document}

\title{Experimental Observation of a Generalized Thouless Pump with a Single Spin}

\author{Wenchao Ma}
\affiliation{CAS Key Laboratory of Microscale Magnetic Resonance and Department of Modern Physics, University of Science and Technology of China (USTC), Hefei 230026, China}
\affiliation{Synergetic Innovation Center of Quantum Information and Quantum Physics, USTC}

\author{Longwen Zhou}
\affiliation{Department of Physics, National University of Singapore, Singapore 117543}

\author{Qi Zhang}
\affiliation{CAS Key Laboratory of Microscale Magnetic Resonance and Department of Modern Physics, University of Science and Technology of China (USTC), Hefei 230026, China}
\affiliation{Synergetic Innovation Center of Quantum Information and Quantum Physics, USTC}

\author{Min Li}
\affiliation{CAS Key Laboratory of Microscale Magnetic Resonance and Department of Modern Physics, University of Science and Technology of China (USTC), Hefei 230026, China}
\affiliation{Synergetic Innovation Center of Quantum Information and Quantum Physics, USTC}

\author{Chunyang Cheng}
\affiliation{CAS Key Laboratory of Microscale Magnetic Resonance and Department of Modern Physics, University of Science and Technology of China (USTC), Hefei 230026, China}
\affiliation{Synergetic Innovation Center of Quantum Information and Quantum Physics, USTC}

\author{Jianpei Geng}
\email{Present address: School of Electronic Science and Applied Physics, Hefei University of Technology, Hefei 230009, China}
\affiliation{CAS Key Laboratory of Microscale Magnetic Resonance and Department of Modern Physics, University of Science and Technology of China (USTC), Hefei 230026, China}

\author{Xing Rong}
\affiliation{CAS Key Laboratory of Microscale Magnetic Resonance and Department of Modern Physics, University of Science and Technology of China (USTC), Hefei 230026, China}
\affiliation{Synergetic Innovation Center of Quantum Information and Quantum Physics, USTC}
\affiliation{Hefei National Laboratory for Physical Sciences at the Microscale, USTC}

\author{Fazhan Shi}
\email{fzshi@ustc.edu.cn}
\affiliation{CAS Key Laboratory of Microscale Magnetic Resonance and Department of Modern Physics, University of Science and Technology of China (USTC), Hefei 230026, China}
\affiliation{Synergetic Innovation Center of Quantum Information and Quantum Physics, USTC}
\affiliation{Hefei National Laboratory for Physical Sciences at the Microscale, USTC}

\author{Jiangbin Gong}
\email{phygj@nus.edu.sg}
\affiliation{Department of Physics, National University of Singapore, Singapore 117543}

\author{Jiangfeng Du}
\email{djf@ustc.edu.cn}
\affiliation{CAS Key Laboratory of Microscale Magnetic Resonance and Department of Modern Physics, University of Science and Technology of China (USTC), Hefei 230026, China}
\affiliation{Synergetic Innovation Center of Quantum Information and Quantum Physics, USTC}
\affiliation{Hefei National Laboratory for Physical Sciences at the Microscale, USTC}


\begin{abstract}
Adiabatic cyclic modulation of a one-dimensional periodic potential will result in quantized charge transport, which is termed the Thouless pump.
In contrast to the original Thouless pump restricted by the topology of the energy band,
here we experimentally observe a generalized Thouless pump that can be extensively and continuously controlled. The extraordinary features of the new pump originate from interband coherence in nonequilibrium initial states, and this fact indicates that a quantum superposition of different eigenstates individually undergoing quantum adiabatic following can also be an important ingredient unavailable in classical physics.
The quantum simulation of this generalized Thouless pump in a two-band insulator is achieved by applying delicate control fields to a single spin in diamond.
The experimental results demonstrate all principal characteristics of the generalized Thouless pump.
Because the pumping in our system is most pronounced around a band-touching point, this work also suggests an alternative means to detect quantum or topological phase transitions. \end{abstract}
\maketitle

In 1983, Thouless discovered that the charge transport across a one-dimensional lattice over an adiabatic cyclic variation of the lattice potential is quantized, equaling to the first Chern number defined over a
Brillouin zone formed by quasimomentum and time~\cite{ThoulessPump1}. This phenomenon, known as the
Thouless pump,
shares the same topological origin as the quantization of Hall conductivity~\cite{TKNN1,TKNN2,TKNN3} and may thus be regarded as a dynamical version of the integer quantum Hall effect~\cite{DQHE1}. In the ensuing years,
the Thouless pump was investigated extensively~\cite{TKNN3}.
Up to now, several single-particle pumping experiments have been implemented in nanoscale devices~\cite{PumpNanoExp1,PumpNanoExp2,PumpNanoExp3,PumpNanoExp4}.
Most recently, the Thouless pump was observed in cold atom systems~\cite{PumpCdAtExp1,PumpCdAtExp2}.
On the application side, the Thouless pump has the potential for realizing novel current standards~\cite{PumpApp1,PumpApp2}, characterizing many-body systems~\cite{PumpInt1,PumpInt2,PumpInt3,PumpInt4,PumpPhoton}, and exploring higher dimensional physics~\cite{PumpHighD1}.

\begin{figure}
\includegraphics[width=1\columnwidth]{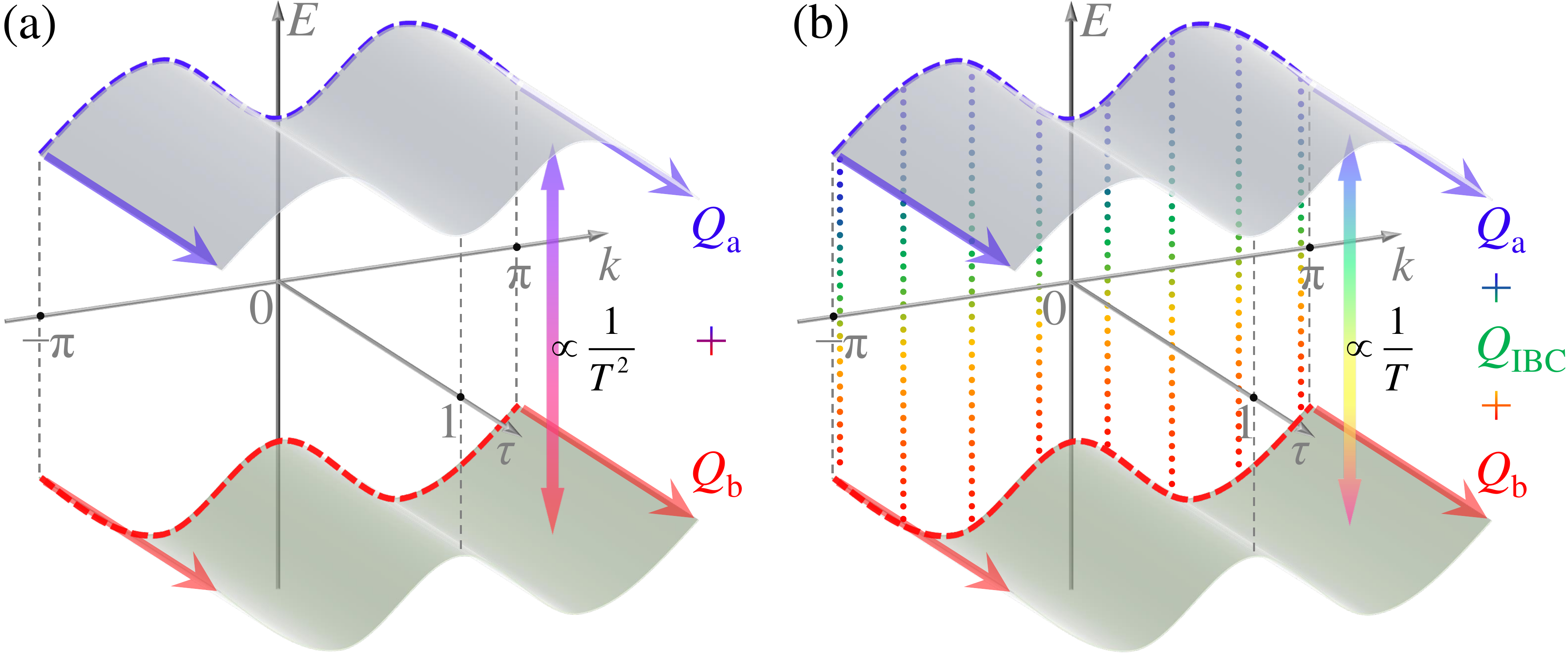}
\caption{
Illustration about how interband coherence in the initial state leads to the generalized Thouless pump of duration $T$.
Band dispersion relations are plotted via energy and quasimomentum variables $E$ vs $k$, and $\tau$ represents the time scaled by $T$.
(a) With an initial state being an incoherent mixture of states from different bands, the nonadiabatic correction to the band populations is of the order of $1/T^2$. The resultant pumping $Q_a + Q_b$ is a weighted sum of the contribution from each band.
(b) With an initial state having interband coherence, the pumping operation induces a population correction of the order of $1/T$, whose effect accumulated over $T$ yields a pumping term $Q_{\text{IBC}}$ that is $T$ independent and continuously tunable.
   }
    \label{theory}
\end{figure}

In Thouless' original proposal and almost all the follow-up studies,
the initial-state quantum coherence between different energy bands, namely, the interband coherence in the initial states, is not taken into account. As a fundamental feature of quantum systems~\cite{Schroedinger,Quantify}, quantum coherence is at the root of a number of fascinating phenomena in chemical physics \cite{CC1,CC2,CC3}, quantum optics~\cite{QO1,QO2,QO3,QO4,QO5}, quantum information~\cite{Chuangbook}, quantum metrology~\cite{Metrology1,Metrology2,Metrology3}, solid-state physics~\cite{Solid1,Solid2}, thermodynamics~\cite{Therm1,Therm2,Therm4,Therm5,Therm6}, magnetic resonance~\cite{MR1,MR2,MR3}, and even biology~\cite{QB1,QB2,QB3}. Therefore, a question naturally arises as to how the pump will behave if the interband coherence resides in the initial state.
A theoretical analysis of this issue is outlined in  Fig.~\ref{theory}, where $Q_a$ and $Q_b$ represent pumping contributed by individual bands, and $Q_{\rm IBC}$ is fueled by interband coherence \cite{PumpWang,PumpZhou}.
In contrast to the conventional Thouless pump where the pumping components $Q_a$ and $Q_b$ are determined by the Berry curvature of each filled band, the generalized Thouless pump is featured by the component $Q_{\rm IBC}$ that can be continuously and extensively controlled.
The main aspects of $Q_{\rm IBC}$ are experimentally investigated in this work.

Consider a one-dimensional two-band insulator subject to time-dependent modulations. Its Hamiltonian in the quasimomentum space is
\begin{eqnarray}
H(k,\tau)&=&\frac{\omega\sin k}{2}\big[\cos\phi(\tau)~\sigma_{x}+\sin\phi(\tau)~\sigma_{y}\big] \nonumber \\
&& + \frac{\delta_1\cos k+\delta_2}{2}~\sigma_{z}.
\label{Hamiltonian}
\end{eqnarray}
Throughout, $\tau=t/T$ is the scaled 
time with $t$ being the real time and $T$ the duration of one pumping cycle, $k\in(-\pi,\pi]$ is the quasimomentum, $\sigma_{x,y,z}$ are the Pauli matrices, and $\hbar$ is set to $1$. The instantaneous spectrum of $H(k,\tau)$ is gapless at $k=0~(k=\pi)$ when and only when $\delta_1=-\delta_2$~($\delta_1=\delta_2$).
	One pumping cycle can be realized by slowly varying $\phi$ from $0$ to $2\pi$.
In a lattice representation, the parameters $\delta_1$ and $\delta_2$ represent the respective bias in the nearest-neighbor hopping strength and energy between two internal states.

For a general initial state with equal populations on quasimomenta $k$ and $-k$ on each band, the pumped amount of charge $Q$ over $N$ adiabatic cycles can be found from the first-order adiabatic perturbation theory (APT)~\cite{Messiah,AdPt3,PumpWang,PumpZhou},
where $Q=N(Q_{\rm TP}+Q_{\rm IBC})+Q_{\rm NG}$, with
\begin{alignat}{1}
 {Q_{{\rm{TP}}}}=& \frac{1}{{2\pi }}\int_{ - \pi }^\pi  {dk\sum\limits_n {{{\left. {{\rho _{nn}}} \right|}_{\tau  = 0}}} \int_0^1 {d\tau }\ {\rm{ }}} {\Omega_{\tau k}^{(n)}}, \\
 {Q_{{\rm{IBC}}}}=& \frac{1}{{2\pi }}\int_{ - \pi }^\pi  dk \sum\limits_{m,n(m < n)}\nonumber \\
 &{\left. {\frac{{2\ {\mathop{\rm Im}\nolimits} \left( {{\rho _{mn}}\left\langle n \right|{\partial _\tau}\left| m \right\rangle } \right)}}{{{E_m} - {E_n}}}} \right|} _{\tau = 0}\int_0^1 {d\tau}~({v_{mm}} - {v_{nn}}),\label{QIBC}
\end{alignat}
and $Q_{\rm NG}$ being a nongeneric term that does not build up with the number of pumping cycles (hence not of interest here)~\cite{SM}. That is, only $Q_{\rm TP}$ and $Q_{\rm IBC}$
represent contributions from pumping over each adiabatic cycle.
In above $m$ and $n$ are band indices, $|m(k,\tau)\rangle$ represent an instantaneous eigenstate of $H(k, \tau)$ with the eigenvalue $E_m(k,\tau)$, $\rho_{mn}(k,\tau)$ and $v_{nm}(k,\tau)$ refer to matrix elements of the density operator and the velocity operator $v(k,\tau)\equiv\partial_k H(k,\tau)$ in representation of $|m(k,\tau)\rangle$, and ${\Omega _{\tau k}^{(n)}}$ is the Berry curvature of the $n$th instantaneous energy band of $H(k,\tau)$.
The component $Q_{\rm TP}$ ($=Q_a+Q_b$ in the case of Fig.~\ref{theory}), representing a weighted integral of the Berry curvature, was found previously by Thouless~\cite{ThoulessPump1}.
The component $Q_{\rm IBC}$, namely, the charge pumping induced by interband coherence in the initial state, is responsible for the generalized Thouless pump and will be our focus here.
Analogous to the conventional Thouless pumping,
$Q_{\rm IBC}$ arises from an accumulation of small nonadiabatic effects over one pumping cycle.
As sketched in Fig.~\ref{theory}, the initial-state interband coherence plays a crucial role in generating the underlying nonadiabatic effects.
The term $Q_{\rm IBC}$ is found to be nontopological and can change continuously.
As indicated by Eq.~(\ref{QIBC}), $Q_{\rm IBC}$ depends on $\langle n|\partial_{\tau}|m\rangle|_{\tau=0}$ and the band gap.
For a pumping parameter $\phi(\tau)$ as in our two-band model depicted in Eq.~(\ref{Hamiltonian}), one has $\langle n|\partial_{\tau}|m\rangle|_{\tau=0} \sim \frac{d\phi(\tau)}{d\tau}|_{\tau=0}$,
which can be controlled by the switching-on rate of a pumping protocol.
The band gap in our model can be altered via tuning $\delta_1/\delta_2$  and $Q_{\rm IBC}$ can even diverge logarithmically as the band gap is tuned to approach zero~\cite{SM}.



\begin{figure}
\includegraphics[width=1\columnwidth]{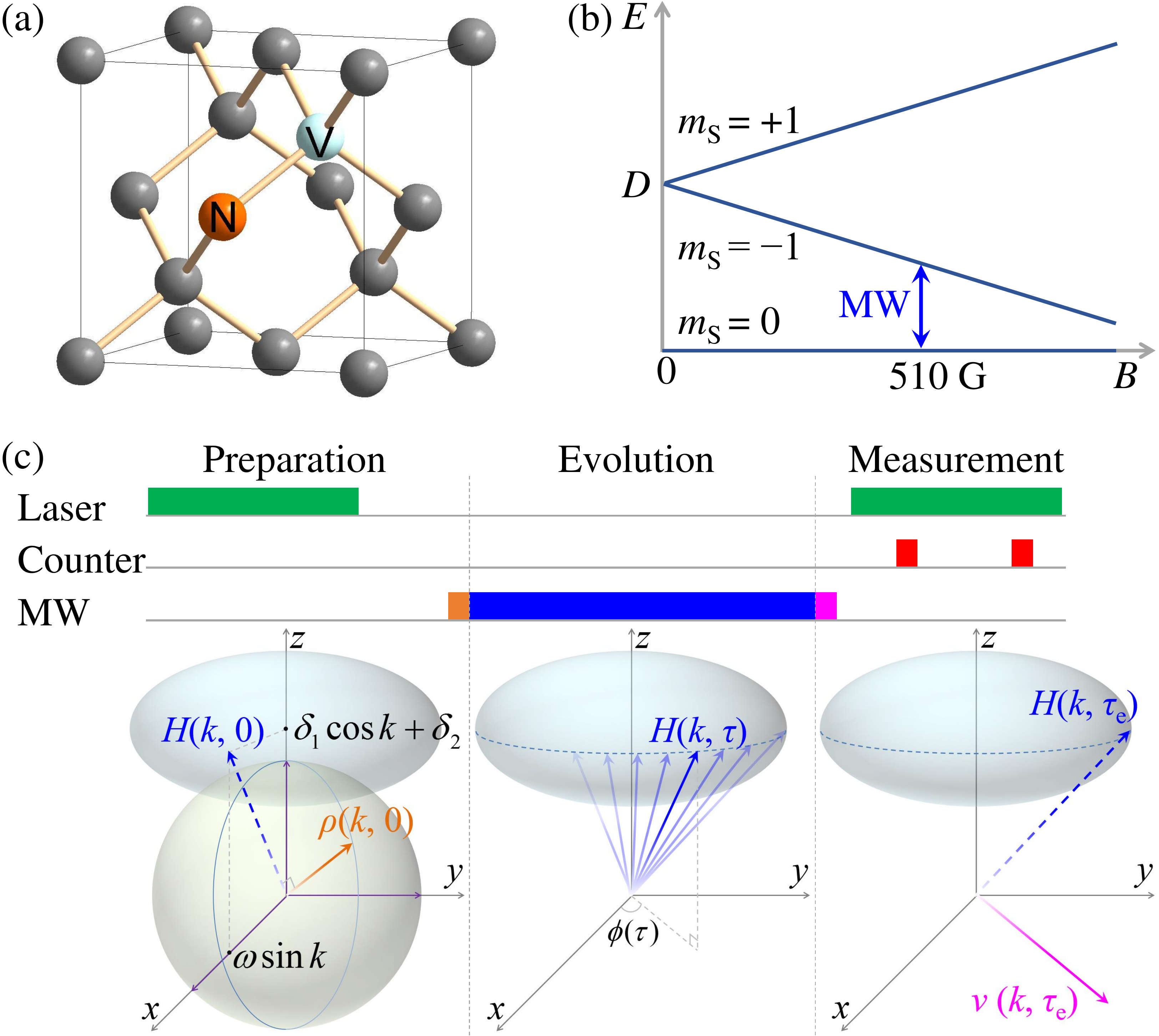}
\caption{Experimental system and method.
   (a) NV center in diamond. 
   (b) Electronic ground state of a negatively charged NV center. The energy splitting depends on the magnetic field which is parallel to the NV axis in this experiment~\cite{NV}. The two levels $\left| {{m_s} = 0} \right\rangle$ and $\left| {{m_s} =  - 1} \right\rangle$ are encoded as a qubit which is manipulated by microwaves (MW).
   (c) Pulse sequence for qubit control and measurement. The ellipsoid surface represents the parameter space and the sphere represents the Bloch sphere.
   }
    \label{structure}
\end{figure}

The generalized Thouless pump can be experimentally realized on a qubit system because the insulator's Hamiltonian in Eq.~(\ref{Hamiltonian}) is also the Hamiltonian of a qubit in a rotating field parametrized by $k$ and $\phi$.
That is, by mapping the two-band insulator's Hamiltonian to that for a qubit in a rotating field, we can experimentally demonstrate the generalized Thouless pump using a single spin \cite{SpHfCherNumExp}.
To highlight the contribution from $Q_{{\rm IBC}}$, in our experiment the initial state is properly designed such that $Q_{\rm TP}=Q_{\rm NG}=0$; i.e., the traditional Thouless pumping and the non-generic term $Q_{\rm NG}$ have no contribution.
To demonstrate the sensitivity of $Q_{{\rm IBC}}$ to the switching-on rate of the pumping protocol, namely, $\frac{d\phi(\tau)}{d\tau}|_{\tau=0}$, we consider a linear ramp $\phi(\tau)=2\pi \tau$ and a quadratic ramp $\phi(\tau)=2\pi\tau^{2}$.
One directly sees that the latter choice with zero switching-on rate will make $Q_{{\rm IBC}}=0$ within the first-order APT.
To demonstrate the dependence of $Q_{{\rm IBC}}$ on the band gap, we implement the Hamiltonian in Eq.~(\ref{Hamiltonian}) with a varying band gap.
A negatively charged nitrogen-vacancy (NV) center in diamond is used in the experiment. 
As shown in Fig.~\ref{structure}(a), the NV center is composed of one substitutional nitrogen atom and an adjacent vacancy~\cite{Doherty,Schirhagl,Prawer,Wrachtrup}.
In our experiment, an external static magnetic field around $510$ G is parallel to the NV symmetry axis. Such magnetic field enables both the NV electron spin and the host $^{14}$N nuclear spin to be polarized by optical excitation~\cite{Jacques,Sar}. As illustrated in Fig.~\ref{structure}(b), microwaves generated by an arbitrary waveform generator drives the transition between the electronic levels $\left| {{m_s} = 0} \right\rangle$ and $\left| {{m_s} =  - 1} \right\rangle$ which compose a qubit, and the level $\left| {{m_s} =  1} \right\rangle$ remains idle due to large detuning~\cite{NV}.
The Hamiltonian of the qubit in the laboratory frame is $H^{\rm{lab}} = \omega_0\sigma_{z}/2 + f(t)\sigma_{x}$,
where the term $f(t)\sigma_{x}$ delineates the effect of the microwave field.
The expectation value of the observable $\sigma_z$ can be read out via fluorescence detection during optical excitation.
All the optical procedure are performed on a home-built confocal microscope, and a solid immersion lens is etched on the diamond above the NV center to enhance the fluorescence collection~\cite{Robledo,Rong}.

In our work, the experiment for different $k$ is performed separately in different runs of experiment. The pulse sequence for each $k$ is sketched in Fig.~\ref{structure}(c).
At first, the qubit is polarized to the state $\rho_0=(\mathbbm{1}+\sigma_z)/2$ by a laser pulse [see the green bar in the preparation section in Fig.~\ref{structure}(c)], and then the initial state $\rho(k,0)$ needs to be prepared.
To optimize $Q_{\text{IBC}}$ in the experiment, the initial density matrix $\rho(k,0)$ at individual values of $k$
is designed such that its associated Bloch vector is perpendicular to the direction of the field
that yields $H(k,0)$.  This choice is again illustrated in Fig.~\ref{structure}(c).
Specifically, $\rho(k,0)$ is chosen as $[\mathbbm{1}+ {\bm{n}(k)} \cdot {\bm{\sigma }}]/2$ with the unit vector $\bm{n}(k)$ along the direction $\left( { - {\delta _1}\cos k  - {\delta _2},0,\omega \sin k } \right)$ except that
$\bm{n}=(0,0,1)$ at a band touching point.
This choice also makes $Q_{\rm{TP}}$ and $Q_{\rm NG }$ vanish~\cite{SM}.
To prepare such an initial state, we apply a resonance microwave pulse with the temporal dependence $f(t) = \omega_1 \cos ( {\omega _0} t +\varphi_{\rm{ini}})$,
where the time $t$ starts from zero, $\omega_1$ is the Rabi frequency, and the initial phase $\varphi_{\rm{ini}}$ is set as $-\pi/2~~(\pi/2)$ if ${\delta _1}\cos k  + {\delta _2}\ge0~~(<0)$.
The duration of the pulse is $t_{\rm{ini}}=\alpha/\omega_1$, where $\alpha$ is the inclination angle of $\bm{n}(k)$. The orange bar in the preparation section in Fig.~\ref{structure}(c) represents this pulse.

Upon initial-state preparation, the qubit is left to evolve under $H(k,\tau)$,
namely, in the presence of a field whose transverse and longitudinal magnitudes are given by $\omega\sin k$ and ${\delta _1}\cos k + {\delta _2}$, respectively.
The field is then rotated around the $z$ axis according to $\phi(\tau)$, with $\phi(\tau)$ understood as the azimuthal angle.
This rotating field is implemented by applying a microwave pulse with $f(t)=\omega \sin k \cos \left[ {({\omega _0} - {\delta _1}\cos k  - {\delta _2})t + \phi (\tau)} + \varphi_{\rm {I}}\right]$, where $t$ starts from zero and the initial phase $\varphi_{\rm {I}}=\omega _0 t_{\rm{ini}}$ is used to match the phase of the first pulse.
In Fig.~\ref{structure}(c) this pulse is depicted by the blue bar in the evolution section.
In a frame with the initial azimuthal angle $\varphi_{\rm {I}}$ and rotating around the $z$ axis with the angular frequency ${\omega}_0 - {\delta}_1 \cos k - {\delta}_2$ relative to the laboratory frame, the bare Hamiltonian $H^{\rm{lab}}$ is transformed, via
the rotation transformation operator ${e^{ - i[({\omega _0} - {\delta _1}\cos k  - {\delta _2})t + \varphi_{\rm {I}}]{\sigma _z}/2}}$,
to our target Hamiltonian $H(k,\tau)$ under the rotating wave approximation.
The parameters adopted in our experiment are $\omega = 2\pi \times 20$ MHz, $\delta_1 = 2\pi \times 10$ MHz, and $\delta_2$ between $0$ to $2\pi \times 20$ MHz.

The evolution governed by $H(k,\tau)$ lasts for some duration $t_{\rm e} \in [0,T]$ (with the corresponding scaled time $\tau_{\rm e} \in [0,1]$), and then the velocity $v(k,\tau_{\rm e})$ needs to be measured.
As shown in Fig.~\ref{structure}(c), the measurement procedure begins with a microwave pulse (the magenta bar).
This pulse is described by $f(t)=\omega_1 \cos ( {\omega _0}t + \varphi_{\rm {I}} + \varphi_{\rm{II}}+ \varphi_{\rm{fin}})$,
with $t$ starting from zero, $\varphi_{\rm{II}}=({\omega _0} - {\delta _1}\cos k  - {\delta _2})t_{\rm e}+\phi (\tau_{\rm{e}})$, and $\varphi_{\rm{fin}} = -\pi/2~~(\pi/2)$ when $\cos k \ge 0~~(<0)$.
The duration of the pulse is $t_{\rm{fin}}=\beta/\omega_1$, where $\beta$ is the inclination angle of the direction of $v$.
This resonant microwave pulse, which steers the direction of $v(k,\tau_{\rm e})$ to the $+z$ direction, is followed by a laser pulse [the right green bar in Fig.~\ref{structure}(c)] together with fluorescence detection.
The fluorescence is collected via two counting windows represented by the two red bars in Fig.~\ref{structure}(c). The former window records the signal while the latter records the reference~\cite{SM}.
The fluorescence collection amounts to the measurement of $\sigma_z$, and the effect combined with the microwave pulse is equivalent to the observation of $v(k,\tau_{\rm e})/ \left\|v\right\|$, where $\left\|v\right\|$ is the spectral norm of $v$.

\begin{figure}\centering
\includegraphics[width=0.95\columnwidth]{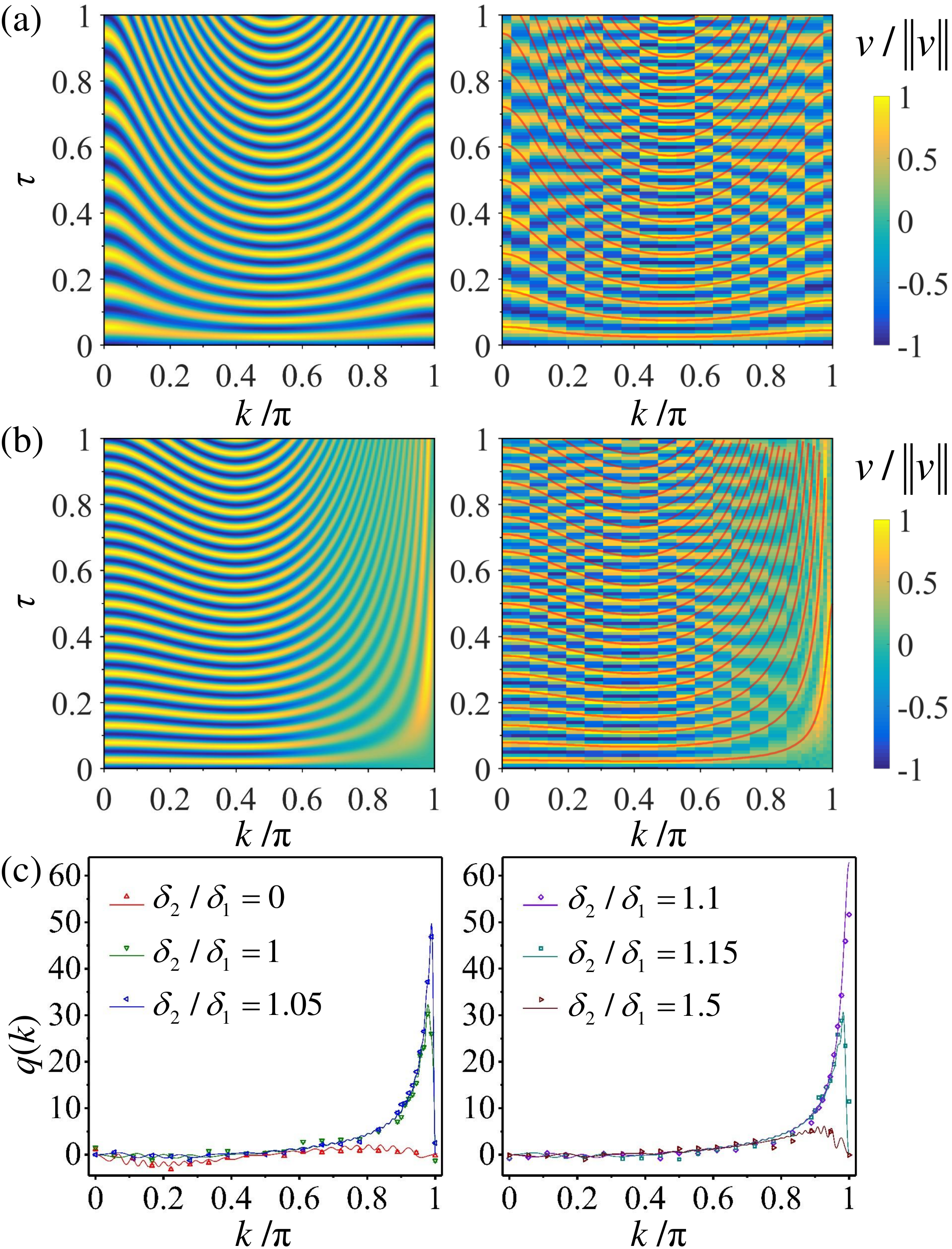}
\caption{Normalized velocity expectation values and pumped charge per each $k$.
   (a),(b) Normalized value $\langle v \rangle / \left\|v\right\|$ vs $k$ and $\tau$ for $\delta_2/\delta_1=0$ and $1$, respectively. Experimental data (calculations based on the Schr\"{o}dinger equation) are on the right (left). The red curves in the experimental figures are guides to the eye to clarify the patterns in the color map. These guidelines are the crest lines in the patterns of the calculated $\langle v \rangle / \left\|v\right\|$.
   (c) Pumped charge $q(k)$ per each synthetic quasimomentum $k$ for several values of $\delta_2/\delta_1$. Symbols (curves) represent the experimental data (the calculation).
   }
    \label{intermediate}
\end{figure}

The above sequence is performed for a series of $\tau_{\rm e}\in [0,1]$, and is iterated at least a hundred thousand times to obtain the expectation value.
One can then get $\langle v(k,\tau) \rangle / \left\|v\right\|$ as a function of $\tau$.
Numerical integration over $\tau$ based on these experimental data, multiplied by $\left\|v\right\|$, yields the experimental value of $q(k)=T\int_{0}^{1}\langle v(k,\tau) \rangle d\tau$, the pumped charge contributed from a certain $k$.
This procedure is repeated for different values of $k \in [0,\pi]$.
Some experimental data with $T=1$ $\rm\mu$s and $\phi(\tau)=2 \pi \tau$ are instantiated in Fig.~\ref{intermediate}. The pattern of the normalized velocity $\langle v(k,\tau) \rangle / \left\|v\right\|$ depends strongly on $\delta_2/\delta_1$, and so does the shape of $q(k)$. In particular, there is a significant charge transport for $\delta_2/\delta_1\approx 1$ and $k\approx \pi$, i.e., near the band touching point.

\begin{figure}\centering
\includegraphics[width=1\columnwidth]{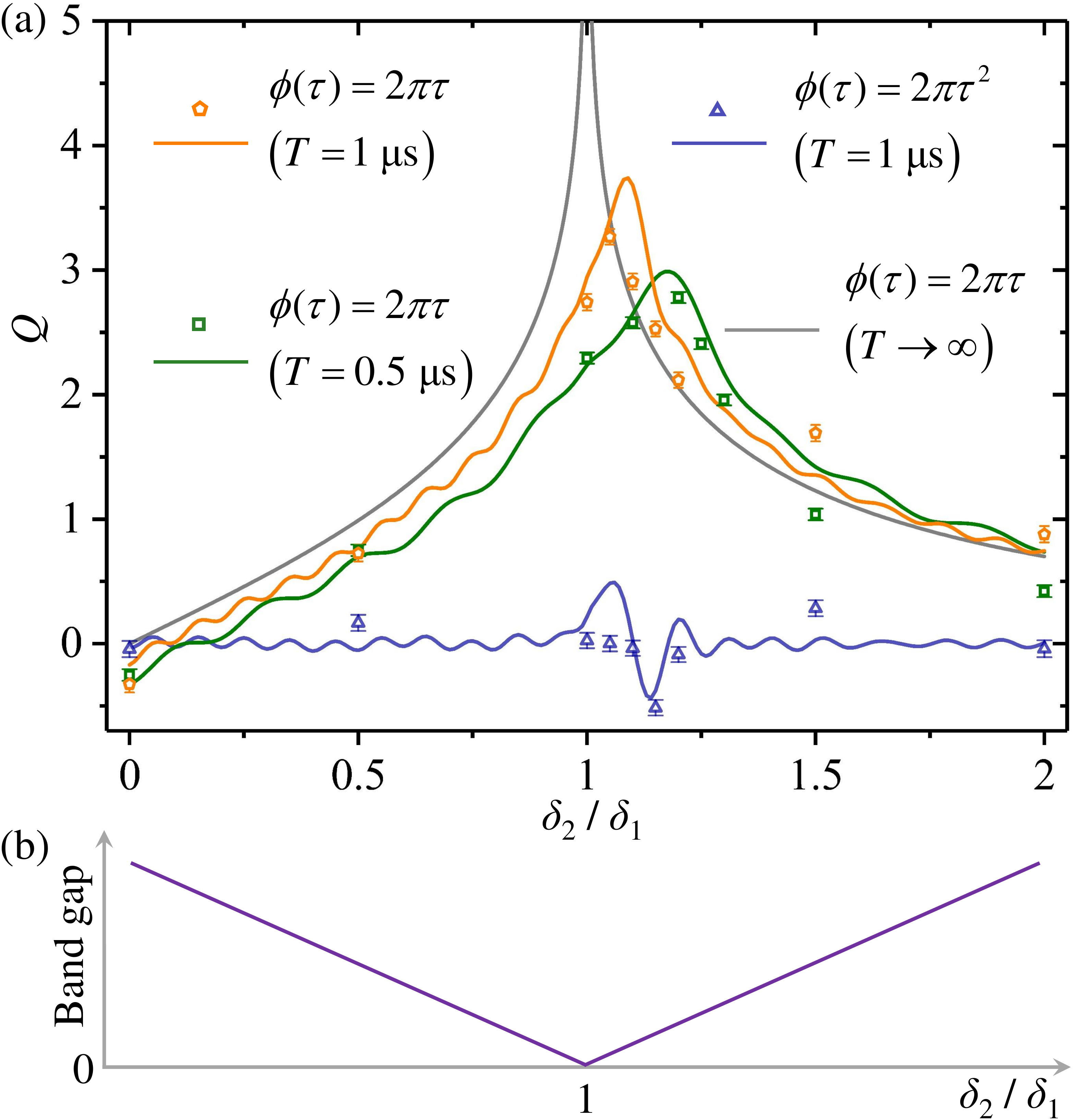}
\caption{Transported charge $Q$ and band gap vs $\delta_2/\delta_1$.
    (a) The symbols and curves represent experimental data and theoretical results, respectively. The orange symbols and curve are for the linear ramp $\phi(\tau)=2 \pi \tau$ with $T=1$ $\rm\mu$s. The green symbols and curve are for $\phi(\tau)=2 \pi \tau$ with $T=0.5$ $\rm\mu$s. The grey theoretical curve corresponds to $\phi(\tau)=2 \pi \tau$ with $T\to\infty$.
    The blue symbols and curve correspond to the parabolic ramp $\phi(\tau)=2 \pi \tau^2$ with $T=1$ $\rm\mu$s. Error bars represent $\pm1$ s.d.
    (b) In the two-band model, the band gap is $|\delta_2-\delta_1|$. 
   }
    \label{final}
\end{figure}

Because of symmetry considerations, it suffices to let our measurements cover half of the first Brillouin zone to extract the pumped charge $Q=\int_{-\pi}^{\pi}q(k)dk/(2\pi)=\int_{0}^{\pi}q(k)dk/\pi$ \cite{SM}.
As illustrated by the orange curve and data points in Fig.~\ref{final}(a), the pumped charge $Q$ first rises and then declines as the parameter $\delta_2/\delta_1$ sweeps from $0$ to $2$.
The parameter $\delta_2/\delta_1$ also determines the band gap as sketched in Fig.~\ref{final}(b).
Though the ramp time $T=1~\rm\mu$s is still not in the true adiabatic limit $T\to\infty$, the pumped charge $Q$ for $T=1~\rm\mu$s as a function of $\delta_2/\delta_1$ bears strong resemblance with the theoretical curve for $T\rightarrow \infty$ obtained using the first-order APT, with their differences well accounted for.
In particular, the theoretical logarithmic divergence of $Q$ as $\delta_2/\delta_1\rightarrow 1$ [see the gray curve in Fig.~\ref{final}(a)] implicitly requires $T\rightarrow \infty$ as the condition to apply the first-order APT.
The actual observed pumping for a finite $T=1$ $\rm\mu$s is thus not expected to shoot to infinity.
In addition, the peak of $Q$
is not precisely at $\delta_2/\delta_1=1$, but has a rightward shift. In this linear ramp case, a non-perturbative theory can be developed~\cite{SM}.
The theoretical shift of the peak of $Q$ as a function of $\delta_2/\delta_1$ is found to be $2\pi/(\delta_1T)$, in good agreement with our observation. This clearly indicates that the observed peak shift is merely a finite-$T$ effect.
For a shorter ramp time $T=0.5$ $\rm\mu$s as depicted by the green curve and data points in Fig.~\ref{final}(a), the pumping peak slightly goes lower again and shifts further away from the exact phase transition point $\delta_2/\delta_1=1$. Overall, the two pumping curves with $T=1$ $\rm\mu$s and $T=0.5$ $\rm\mu$s have a remarkable overlap with each other, thus supporting that to the zeroth order of $1/T$, the outcome of the generalized Thouless pump is independent of $T$.  We next investigate another pumping protocol $\phi(\tau)=2 \pi \tau^2$ with $T=1$ $\rm\mu$s. The initial switching-on rate of this pumping protocol now vanishes. In this case, we observe negligible pumping, as evidenced by the blue curve and data points in Fig.~\ref{final}(a).
The results for the two different protocols confirm that the generalized Thouless pump can be extensively tuned by varying the switching-on rate of a pumping protocol. Finally, one may note the differences between experimental results and the simulation results [orange, green, and blue solid curves in Fig.~\ref{final}(a)] based solely on time-dependent Schr\"odinger equations. The experimental errors are mainly due to the imperfection of the microwave pulses. Nevertheless, in the presence of the experimental errors, our experimental results have demonstrated all principal features of the generalized Thouless pump.
In conclusion, by incorporating interband coherence into the initial state as a powerful quantum resource, we are able to go beyond the traditional Thouless
pump. Using a single spin in diamond,  we have experimentally demonstrated a novel type of quantum adiabatic pump, which is extensively and continuously tunable by varying the switching-on rate of a pumping protocol.
The tunability of our generalized Thouless pump is reminiscent of the famous Archimedes screw, where water is pumped via rotating a screw-shaped blade in a cylinder and the amount of pumped water can also be changed continuously~\cite{Archimedes1,Archimedes2,Archimedes3}.  Furthermore, because the coherence-based pumping in our system is most pronounced around a band-touching point, it may provide an alternative
means for the detection of band touching and hence quantum or topological phase transition points.
Our work thus  enriches the physics of adiabatic pump and coherence-based quantum control.

\begin{acknowledgements}
The authors at University of Science and Technology of China are supported by the National Natural Science Foundation of China (Grants No.~81788101, No.~11227901, No.~31470835, No.~91636217, and No.~11722544), the CAS (Grants No.~GJJSTD20170001, No.~QYZDY-SSW-SLH004, No.~QYZDB-SSW-SLH005, and No.~YIPA2015370), the 973 Program (Grants No.~2013CB921800, No.~2016YFA0502400, and No.~2016YFB0501603), the CEBioM, and the Fundamental Research Funds for the Central Universities (WK2340000064). The authors at National University of Singapore are supported by the Singapore NRF Grant No.~NRF-NRFI2017-04 (WBS No.~R-144-000-378-281) and by the Singapore Ministry of Education Academic Research Fund Tier I (WBS No.~R-144-000-353-112).

W.~M., L.~Z., and Q.~Z. contributed equally to this work.
\end{acknowledgements}

\onecolumngrid
\vspace{1.5cm}

\begin{center}
\textbf{\large Supplementary Material}
\end{center}

\setcounter{figure}{0}
\setcounter{equation}{0}
\makeatletter
\renewcommand{\thefigure}{S\@arabic\c@figure}
\renewcommand{\theequation}{S\@arabic\c@equation}
\renewcommand{\bibnumfmt}[1]{[S#1]}
\renewcommand{\citenumfont}[1]{S#1}


\section{1. Theory}
\subsection{1.1 Adiabatic charge pumping with nonequilibrium initial states}
In this supplementary note, we derive Eqs.~(2) and (3) in the main text, which describe the particle pumping over an adiabatic cycle
in the generalized Thouless pump for initial states with interband coherence. Throughout this note, we take $\hbar=1$.

We start with the time-dependent Schr\"{o}dinger equation
\begin{equation}
i\frac{d}{d\tau}|\Psi(k,\tau)\rangle=TH(k,\tau)|\Psi(k,\tau)\rangle,\label{eq:Seq}
\end{equation}
where $\tau=t/T\in[0,1]$ is the scaled time with $t$ being the real time and $T$ the duration of the evolution, $k$ represents some other time-independent parameters of the system, $|\Psi(k,\tau)\rangle$ represents the state of the system at the scaled time $\tau$, and $H(k,\tau)$ is the system's Hamiltonian which depends on time through some parameter such as $\phi(\tau)$.
In this study, we consider the class of quantum systems whose Hamiltonian $H(k,\tau)$ admits a discrete instantaneous energy spectrum $\{E_{n}(k,\tau)\}$ with eigenstates $\{|n(k,\tau)\rangle\}$, such that
$$H(k,\tau)|n(k,\tau)\rangle=E_{n}(k,\tau)|n(k,\tau)\rangle,$$
with $n$ being the energy level index.
At the start of the evolution ($\tau=0$), the initial state of the system $|\Psi(k,0)\rangle$ can be written in the basis $\{|n(k,0)\rangle\}$ as
\begin{equation}
|\Psi(k,0)\rangle=\sum_{n}c_{n}(k,0)|n(k,0)\rangle,
\label{eq:Ini-State}
\end{equation}
with the amplitude $c_{n}(k,0)=\langle n(k,0)|\Psi(k,0)\rangle$.
At a later time $\tau$, the state of the system can be written in the basis $\{|n(k,\tau)\rangle\}$ as
\begin{equation}
|\Psi(k,\tau)\rangle=\sum_{n}e^{-i\Theta_{n}(k,\tau)}c_{n}(k,\tau)|n(k,\tau)\rangle,\label{eq:State}
\end{equation}
where $\Theta_{n}(k,\tau)=T\int_{0}^{\tau}E_{n}(k,\tau')d\tau'$ is the dynamical phase. The Schr\"{o}dinger equation in Eq.~(\ref{eq:Seq}) is solved if all $\{c_{n}(k,\tau)\}$ are found at each $\tau$.

In quasiadiabatic evolutions, $\phi$ varies slowly in time, so that
$\frac{d\phi}{dt}=\frac{1}{T}\frac{d\phi(\tau)}{d\tau}$ is much smaller
than any energy gap of the instantaneous Hamiltonian $H(k,\tau)$.
In this case, $c_{n}(k,\tau)$ can be expressed as a series of $1/T$
through adiabatic perturbation theory \cite{APT1}. Keeping terms
up to ${\cal O}(1/T)$, we get
\begin{equation}
c_{n}(k,\tau)=c_{n}(k,0)+\frac{1}{T}\sum_{m\neq n}c_{m}(k,0)\left.\left[\frac{i\langle n(k,\tau')|\partial_{\tau}m(k,\tau')\rangle}{E_{n}(k,\tau')-E_{m}(k,\tau')}e^{i\Theta_{nm}(k,\tau')}\right]\right|_{\tau'=0}^{\tau'=\tau},\label{eq:Amp}
\end{equation}
where $\Theta_{nm}(k,\tau)\equiv\Theta_{n}(k,\tau)-\Theta_{m}(k,\tau)$ is the dynamical phase difference \cite{APT2}. The above equation can also be expressed by the element of the density matrix, namely,
\begin{equation}\label{element}
\begin{aligned}
\rho_{mn}(k,\tau)=\rho_{mn}(k,0)+&\frac{1}{T}\sum_{\ell\neq n}\rho_{m\ell}(k,0)\left.\left[\frac{i\langle \ell(k,\tau')|\partial_{\tau}n(k,\tau')\rangle}{E_{n}(k,\tau')-E_{\ell}(k,\tau')}e^{i\Theta_{\ell n}(k,\tau')}\right]\right|_{\tau'=0}^{\tau'=\tau}\\
+&\frac{1}{T}\sum_{\ell\neq m}\rho_{\ell n}(k,0)\left.\left[\frac{i\langle m(k,\tau')|\partial_{\tau}\ell(k,\tau')\rangle}{E_{m}(k,\tau')-E_{\ell}(k,\tau')}e^{i\Theta_{m\ell}(k,\tau')}\right]\right|_{\tau'=0}^{\tau'=\tau},
\end{aligned}
\end{equation}
Note that in writing down the expressions in Eqs.~(\ref{eq:Amp}) and (\ref{element}), we have taken the parallel transport gauge convention. This means to choose the phase for the basis state $|n(k,\tau)\rangle$, in order to make $\langle n(k,\tau)|\partial_{\tau}|n(k,\tau)\rangle=0$ for all $n$ at any $\tau\in(0,1)$.

In Thouless' setup of adiabatic charge transport~\cite{ThouPump1}, $H(k,\tau)$ describes noninteracting electrons in a one-dimensional lattice modulated by a slowly varying time-dependent potential, which is periodic in both space and time. The particle transported across the system over an adiabatic driving cycle (i.e., $\tau:0\rightarrow1$ here) is given by
\begin{equation}
Q=\frac{1}{2\pi}\int_{-\pi}^{\pi}dk\int_{0}^{1}d\tau T\langle v(k,\tau) \rangle,\label{eq:Q}
\end{equation}
where $k\in(-\pi,\pi]$ is the quasimomentum~(with lattice constant $a=1$), $v(k,\tau)\equiv\partial_{k}H(k,\tau)$ represents the group velocity operator, and $\langle v(k,\tau) \rangle\equiv{\rm{tr}}[\rho(k,\tau) v(k,\tau)]$ represents the expectation value of $v(k,\tau)$.

In the following, we give the detailed derivation of the charge pumping $Q$ discussed in the main text. Let's first introduce a set of compact notations as
\begin{align}
D_{mn}(k,\tau) & \equiv E_{m}(k,\tau)-E_{n}(k,\tau),\\
v_{mn}(k,\tau) & \equiv\langle m(k,\tau)|v(k,\tau)|n(k,\tau)\rangle,\\
M_{mn}(k,\tau) & \equiv i\langle m(k,\tau)|\partial_{\tau}|n(k,\tau)\rangle.
\end{align}
In terms of these notations, we can organize the density matrix components in Eq.~(\ref{element}) as
\begin{alignat}{1}
\varrho_{{\rm I}} & \equiv\rho_{mn}(k,0),\label{eq:rhoI}\\
\varrho_{{\rm II}} & \equiv\frac{1}{T}\sum_{\ell\neq n}\rho_{m\ell}(k,0)\left.\left[\frac{M_{\ell n}(k,\tau')}{D_{n\ell}(k,\tau')}e^{i\Theta_{\ell n}(k,\tau')}\right]\right|_{\tau'=0}^{\tau'=\tau},\label{eq:rhoII}\\
\varrho_{{\rm III}} & \equiv\frac{1}{T}\sum_{\ell\neq m}\rho_{\ell n}(k,0)\left.\left[\frac{M_{m\ell}(k,\tau')}{D_{m\ell}(k,\tau')}e^{i\Theta_{m\ell}(k,\tau')}\right]\right|_{\tau'=0}^{\tau'=\tau}.\label{eq:rhoIII}
\end{alignat}
Correspondingly, we will also decompose the charge pumping $Q$ into three parts as
\begin{equation}
Q = Q_{{\rm NG}}+Q_{{\rm II}}+Q_{{\rm III}}.
\end{equation}
Explicit expressions for these components will be derived in the following subsections.

\subsubsection{1.1.1 Derivation of $Q_{{\rm NG}}$}
The contribution of $\varrho_{{\rm I}}$ to $Q$ is denoted by
$Q_{{\rm NG}}$. From Eqs.~(\ref{eq:Q}) and (\ref{eq:rhoI}), we find
\begin{equation}
Q_{{\rm NG}} = \frac{1}{2\pi}\int_{-\pi}^{\pi}dk\sum_{m,n}\rho_{nm}(k,0)\int_{0}^{1} T d\tau v_{mn}(k,\tau)e^{i\Theta_{mn}(k,\tau)}.\label{eq:QTS1}
\end{equation}
In the case of $m=n$, the integral becomes $\frac{1}{2\pi}\int_{-\pi}^{\pi}dk\sum_{n}\rho_{nn}(k,0)\int_{0}^{1} T d\tau v_{nn}(k,\tau)$.
If there is symmetry breaking in $k$-space (e.g., a population imbalance with respect to $k$), this term could make a contribution to the transport of order $T$, which may become very large in the adiabatic limit ($T\rightarrow\infty$). But this contribution is not due to pumping. To remove this irrelevant term, we will assume
\begin{equation}
\frac{1}{2\pi}\int_{-\pi}^{\pi}dk\sum_{n}\rho_{nn}(k,0)\int_{0}^{1} T d\tau v_{nn}(k,\tau)=0.\label{eq:Assumption1}
\end{equation}
Practically this can be achieved if, e.g., $v_{nn}(k,\tau)$ and $\rho_{nn}(k,0)$ have opposite parities as functions of $k$. In our experiment, we studied a two-band system and choose the initial state to equally populate the two bands, i.e., $\rho_{11}(k,0)=\rho_{22}(k,0)$ for all $k$. Since the group velocities of the two bands satisfy $v_{11}(k,\tau)=-v_{22}(k,\tau)$, we will always have $\sum_{n=1}^{2}\rho_{nn}(k,0)v_{nn}(k,\tau)=0$ in our experimental situation, and therefore the assumption (\ref{eq:Assumption1}) is always justified.

Under the assumption (\ref{eq:Assumption1}), Eq.~(\ref{eq:QTS1}) simplifies to
\begin{equation}
Q_{{\rm NG}} = \frac{1}{2\pi}\int_{-\pi}^{\pi}dk\sum_{m,n,m\neq n}\rho_{nm}(k,0)\int_{0}^{1} T d\tau v_{mn}(k,\tau)e^{i\Theta_{mn}(k,\tau)}.
\end{equation}
Performing an integration by parts over the dynamical phase exponent
$e^{i\Theta_{mn}(k,\tau)}$, we find
\begin{alignat}{1}
Q_{{\rm NG}} & = \frac{1}{2\pi}\int_{-\pi}^{\pi}dk\sum_{m,n,m\neq n}\rho_{nm}(k,0)\int_{0}^{1}\frac{v_{mn}(k,\tau)}{i D_{mn}(k,\tau)}de^{i\Theta_{mn}(k,\tau)}\nonumber \\
& =\frac{1}{2\pi}\int_{-\pi}^{\pi}dk\sum_{m,n,m\neq n}\rho_{nm}(k,0)\left.\left[\frac{v_{mn}(k,\tau)}{iD_{mn}(k,\tau)}e^{i\Theta_{mn}(k,\tau)}\right]\right|_{\tau=0}^{\tau=1}+{\cal O}\left(\frac{1}{T}\right).\label{eq:QTS2}
\end{alignat}
Since in the adiabatic limit ($T\rightarrow\infty$), the phase factor $\Theta_{mn}(k,\tau)$ is oscillating fast with respect to $k$, its average over $k$ will tend to vanish. This may also be seen by performing another integration by parts over $e^{i\Theta_{mn}(k,\tau)}$, which will generate a term $1/\partial_{k}\Theta_{mn}(k,\tau)\propto1/T$.
So in the adiabatic limit ($T\rightarrow\infty$), Eq.~(\ref{eq:QTS2}) further
reduces to
\begin{equation}
Q_{{\rm NG}} = \frac{1}{2\pi}\int_{-\pi}^{\pi}dk\sum_{m,n,m<n}\frac{2{\rm Im}\left[\rho_{mn}(k,0)v_{nm}(k,0)\right]}{E_{m}(k,0)-E_{n}(k,0)}.\label{eq:Qts}
\end{equation}
This term is highly non-generic. It contains the memory of the state to the initial condition and is time-independent~(since it is only evaluated at $\tau=0$).  Furthermore, $Q_{{\rm NG}}$ will not accumulate with the increasing of the number of pumping cycles, and thus of secondary importance in long time dynamics. In our experiment, the initial states we prepared satisfy ${\rm Im}\left[\rho_{mn}(k,0)v_{nm}(k,0)\right]=0$, and therefore make $Q_{{\rm NG}}$ vanish as mentioned in the main text.
\subsubsection{1.1.2 Derivation of $Q_{{\rm TP}}$ and $Q_{{\rm IBC}}$}
Plugging Eq.~(\ref{eq:rhoII}) into Eq.~(\ref{eq:Q}), we find
\begin{equation}
Q_{{\rm II}} = \frac{1}{2\pi}\int_{-\pi}^{\pi}dk\sum_{m,n}\sum_{\ell\neq m}\rho_{n\ell}(k,0)\int_{0}^{1}d\tau v_{mn}(k,\tau)\left.\left[\frac{M_{\ell m}(k,\tau')}{D_{m\ell}(k,\tau')}e^{i\Theta_{\ell m}(k,\tau')}\right]\right|_{\tau'=0}^{\tau'=\tau}e^{i\Theta_{mn}(k,\tau)}.\label{eq:QII1}
\end{equation}
When $m=n$, the factor $\frac{M_{\ell m}(k,\tau')}{D_{m\ell}(k,\tau')}e^{i\Theta_{\ell m}(k,\tau')}$ has contribution to $Q_{{\rm II}}$ in the adiabatic limit only at $\tau'=0$, where $e^{i\Theta_{\ell m}(k,\tau')}=1$. When $m\neq n$, the factor $\frac{M_{\ell m}(k,\tau')}{D_{m\ell}(k,\tau')}e^{i\Theta_{\ell m}(k,\tau')}$ has contribution to $Q_{{\rm II}}$ in the adiabatic limit only at $\tau'=\tau$ with $\ell=n$. These can be derived by performing integration by parts over the dynamical phase exponent $e^{i\Theta_{\ell m}(k,\tau)}$, and arguments parallel to what we have in used the last subsection to obtain Eq.~(\ref{eq:QTS2}). Taking the adiabatic limit ($T\rightarrow\infty$) and collecting all non-vanishing terms of Eq.~(\ref{eq:QII1}), we obtain
\begin{alignat}{1}
Q_{{\rm II}} & =\frac{1}{2\pi}\int_{-\pi}^{\pi}dk\sum_{m,n,m\neq n}\rho_{nm}(k,0)\left.\left[\frac{M_{mn}(k,\tau)}{D_{mn}(k,\tau)}\right]\right|_{\tau=0}\int_{0}^{1}d\tau v_{nn}(k,\tau)\nonumber \\
& +\frac{1}{2\pi}\int_{-\pi}^{\pi}dk\sum_{m,n,m\neq n}\rho_{nn}(k,0)\int_{0}^{1}dsv_{mn}(k,\tau)\frac{M_{nm}(k,\tau)}{D_{mn}(k,\tau)}.\label{eq:QII2}
\end{alignat}
Similarly, plugging Eq.~(\ref{eq:rhoIII}) into Eq.~(\ref{eq:Q}) yields
\begin{equation}
Q_{{\rm III}}=\frac{1}{2\pi}\int_{-\pi}^{\pi}dk\int_{0}^{1}d\tau\sum_{m,n}\sum_{\ell\neq n}v_{mn}(k,\tau)\rho_{\ell m}(k,0)\left.\left[\frac{M_{n\ell}(k,\tau')}{D_{n\ell}(k,\tau')}e^{i\Theta_{n\ell}(k,\tau')}\right]\right|_{\tau'=0}^{\tau'=\tau}e^{i\Theta_{mn}(k,\tau)}.\label{eq:QIII1}
\end{equation}
Parallel to the previous analysis, this expression reduces in the adiabatic limit to
\begin{alignat}{1}
Q_{{\rm III}} & =\frac{1}{2\pi}\int_{-\pi}^{\pi}dk\sum_{m,n,m\neq n}\rho_{mn}(k,0)\left.\left[\frac{M_{nm}(k,\tau)}{D_{mn}(k,\tau)}\right]\right|_{\tau=0}\int_{0}^{1}d\tau v_{nn}(k,\tau)\nonumber \\
& +\frac{1}{2\pi}\int_{-\pi}^{\pi}dk\sum_{m,n,m\neq n}\rho_{nn}(k,0)\int_{0}^{1}d\tau v_{nm}(k,\tau)\frac{M_{mn}(k,\tau)}{D_{mn}(k,\tau)}.\label{eq:QIII2}
\end{alignat}
It is not hard to see that $Q_{{\rm III}}=Q_{{\rm II}}^{*}$. Thus we can collect them together and recombine relevant terms to obtain $Q_{{\rm II}}+Q_{{\rm III}}=Q_{{\rm TP}}+Q_{{\rm IBC}}$.

To summarize, the total charge pumping $Q$ can be expressed as a summation of three components as discussed in the main text:
\begin{equation}
Q = Q_{{\rm NG}}+Q_{{\rm II}}+Q_{{\rm III}}
  = Q_{{\rm NG}}+Q_{{\rm TP}}+Q_{{\rm IBC}}.\label{eq:PumpQ}
\end{equation}
On the right hand side of Eq.~(\ref{eq:PumpQ}), the term $Q_{{\rm NG}}$ has been discussed in the last section. The $Q_{{\rm TP}}$ has the following expression:
\begin{equation}
Q_{{\rm TP}}=\frac{1}{2\pi}\int_{-\pi}^{\pi}dk\sum_{n}\rho_{nn}(k,0)\int_{0}^{1}d\tau\Omega_{\tau k}^{(n)},\label{eq:Qeq}
\end{equation}
where $\left.\rho_{nn}(k,\tau)\right|_{\tau=0}=|c_{n}(k,0)|^{2}$ is the initial population at the quasimomentum $k$ on the Bloch band $n$, and $\Omega_{\tau k}^{(n)}=i\langle\partial_{\tau}n(k,\tau)|\partial_{k}n(k,\tau)\rangle+{\rm c.c.}$ represents the Berry curvature. Therefore, $Q_{{\rm TP}}$ is given by an integral of the Berry curvature weighted by initial Bloch band populations. It has been found in Thouless' original work~\cite{ThouPump1}, but has nothing to do with interband coherence in the initial state.

The third term, $Q_{{\rm IBC}}$, is the focus of our experimental study. It is given by
\begin{equation}
Q_{{\rm IBC}}=\frac{1}{2\pi}\int_{-\pi}^{\pi}dk\sum_{m,n,m<n}\left.\frac{2{\rm Im}\left[\rho_{mn}(k,\tau)\langle n(k,\tau)|\partial_{\tau}|m(k,\tau)\rangle\right]}{E_{m}(k,\tau)-E_{n}(k,\tau)}\right|_{\tau=0}\int_{0}^{1}d\tau[v_{mm}(k,\tau)-v_{nn}(k,\tau)].\label{eq:Qibc}
\end{equation}
Through its dependence on $\rho_{mn}(k,\tau)$ at $\tau=0$ for $m\neq n$, we see that $Q_{{\rm IBC}}$ is originated from interband coherence in the initial state. As can be seen from Eq.~(\ref{eq:Amp}), such an initial-state coherence could induce a correction to interband population transfer of the order of $1/T$. The accumulation of this nonadiabatic effect over a long time duration $T$ finally makes $Q_{{\rm IBC}}$ an important component of the total charge pumping. Moreover, $Q_{{\rm IBC}}$ depends on the term $\langle n|\partial_{\tau}|m\rangle=\frac{d\phi}{d\tau}\langle n|\partial_{\phi}|m\rangle$ evaluated at $\tau=0$, and is therefore sensitive to the switching-on behavior of a pumping protocol. In our experiment, we considered two different adiabatic protocols. The first protocol, $\phi(\tau)=2\pi\tau$, is linear in $\tau$ with a constant rate $\frac{d\phi}{d\tau}=2\pi$. The second protocol, $\phi(\tau)=2\pi\tau^{2}$, is quadratic in $\tau$. It has a rate $\frac{d\phi}{d\tau}=4\pi\tau$, vanishing at $\tau=0$. So for the second protocol, one has $Q_{{\rm IBC}}=0$ within the first-order adiabatic perturbation theory.
\subsubsection{1.1.3 Pumping over $N$ adiabatic cycles}
If the pump is operated over $N$ adiabatic cycles, the non-generic part of charge pumping is still given by Eq.~(\ref{eq:Qts}) since its right-hand side is independent of $\tau$. Due to the periodicity of $\Omega^{(n)}_{\tau k}$ in $\tau$, i.e., $\Omega^{(n)}_{\tau+1 k}=\Omega^{(n)}_{\tau k}$, we have $\int_{0}^{N}d\tau\Omega_{\tau k}^{(n)}=N\int_{0}^{1}d\tau\Omega_{\tau k}^{(n)}$ and therefore the relevant pumping component over $N$ adiabatic cycle is $N Q_{{\rm TP}}$, where $Q_{{\rm TP}}$ is the component over one adiabatic cycle. Similarly, the velocity operator $v(k,\tau)$ is also a periodic function of $\tau$ with period $1$, thus $\int_{0}^{N}d\tau[v_{mm}(k,\tau)-v_{nn}(k,\tau)]=N\int_{0}^{1}d\tau[v_{mm}(k,\tau)-v_{nn}(k,\tau)]$ and therefore the relevant pumping component over $N$ adiabatic cycle is $N Q_{{\rm IBC}}$, where $Q_{{\rm IBC}}$ is the component over one adiabatic cycle. Collecting all these together, we conclude that the charge pumping over $N$ adiabatic cycles is $N(Q_{{\rm TP}}+Q_{{\rm IBC}})+Q_{{\rm NG}}$ as discussed in the main text. Here $N$ is a positive integer.


\subsection{1.2 Model for the experiment}
In our experiment, we map a one-dimensional two-band insulator model onto a single qubit subject to a time-dependent driving field. Explicitly,
the qubit Hamiltonian is given by
\begin{equation}
H(k,\tau)=\frac{\omega\sin (k)}{2}\big\{\cos[\phi(\tau)]~\sigma_{x}+\sin[\phi(\tau)]~\sigma_{y}\big\} + \frac{\delta_1\cos(k)+\delta_2}{2}~\sigma_{z},
\label{eq:Hqubit}
\end{equation}
Its corresponding lattice Hamiltonian is
\begin{equation}
H(\tau ) = \sum\limits_k {H(k,\tau )\left| k \right\rangle \left\langle k \right|},
\end{equation}
where $k$ represents the quasimomentum. In the position representation, the Hamiltonian is given by
\begin{equation}
H(\tau)=\frac{1}{2}\sum_{j}\left\{\left| j \right\rangle\frac{\delta_{1}\sigma_{z}-i\omega\cos[\phi(\tau)]\sigma_{x}-i\omega\sin[\phi(\tau)]\sigma_{y}}{2}\left\langle j+1 \right|+{\rm h.c.}\right\}+\frac{1}{2}\sum_{j}\left| j \right\rangle\left(\delta_{2}\sigma_{z}\right)\left\langle j \right|,
\end{equation}
where $\left| j \right\rangle$ represents the lattice site basis, $\delta_{1}$ represents an energy bias in the hopping of spin up and down particles, and $\delta_{2}$ represents an energy bias between spin up and down particles in the same unit cell. The driving field modulates the spin orientation of the particle on $xy$ plane during its hopping between nearest neighbor sites.

In the experiment, we conduct the charge pumping along the synthetic dimension $k$ on a qubit with the Hamiltonian $H(k,\tau)$.
To single out the contribution of $Q_{{\rm IBC}}$ from the total charge pumping $Q$ in an adiabatic cycle, the initial state (at $\tau=0$) is chosen to have equal populations on the two levels of $H(k,0)$ at each $k$.
From Eq.~(\ref{eq:Qeq}), we observe that $Q_{{\rm TP}}=0$ for such an initial state, since the Berry curvature satisfies $\sum_{n}\Omega_{\tau k}^{(n)}=0$ at each point $(k,\tau)$ in the parameter space.
Actually, in this specific model, one has $Q_{{\rm TP}}=0$ even if the initial populations on the two bands are not equal because the first Chern numbers vanish, i.e., the integral of Berry curvature $\frac{1}{2\pi}\int_{-\pi}^{\pi}dk\int_{0}^{1}d\tau\Omega_{\tau k}^{(n)}$ is zero. Despite this particularity of our two-band model, preparing the initial state with equal populations on all bands at each $k$ is an effective method to eliminate $Q_{{\rm TP}}$. It should also be noted that $Q_{{\rm IBC}}$ given by Eq.~(\ref{eq:Qibc}) is a natural consequence of the initial-state interband coherence and is not restricted to the model adopted in this work.

The eigenenergies and eigenstates of $H(k,\tau)$ are
\begin{equation}
E_{\pm}(k)=\pm\frac{1}{2}\sqrt{\nu^{2}+\delta^{2}},\qquad
|+(k,\tau)\rangle=\frac{1}{\sqrt{2}}\begin{bmatrix}\sqrt{1+\frac{\delta}{\Delta}}\\
{\rm sgn}(\frac{\nu}{\Delta})e^{i\phi}\sqrt{1-\frac{\delta}{\Delta}}
\end{bmatrix},\qquad|-(k,\tau)\rangle=\frac{1}{\sqrt{2}}\begin{bmatrix}{\rm sgn}(\frac{\nu}{\Delta})\sqrt{1-\frac{\delta}{\Delta}}\\
-e^{i\phi}\sqrt{1+\frac{\delta}{\Delta}}
\end{bmatrix},
\label{eigen1}
\end{equation}
with
\begin{equation}
\Delta\equiv E_{+}(k)-E_{-}(k)=\sqrt{\nu^{2}+\delta^{2}},\qquad\nu\equiv\omega\sin(k),\qquad\delta\equiv\delta_{1}\cos(k)+\delta_{2}.\label{eq:E-nu-del}
\end{equation}
Here $\Delta$ represents the level spacing, $\nu$ is the magnitude of the transverse field, and $\delta$ is the magnitude of the longitudinal field. The ${\rm sgn}$ function in Eq.~(\ref{eigen1}) reflects a gauge choice. Note also that in our model $E_{\pm}(k)$ and $\Delta$ are independent of time.
Consider next the following initial state for each $k$,
\begin{equation}
|\Psi(k,0)\rangle=\frac{1}{\sqrt{2}}[|+(k,0)\rangle+|-(k,0)\rangle]=\frac{1}{2}\begin{bmatrix}\sqrt{1+\frac{\delta}{\Delta}}+{\rm sgn}(\frac{\nu}{\Delta})\sqrt{1-\frac{\delta}{\Delta}}\\
{\rm sgn}(\frac{\nu}{\Delta})\sqrt{1-\frac{\delta}{\Delta}}-\sqrt{1+\frac{\delta}{\Delta}}
\end{bmatrix},\label{eq:StateIni}
\end{equation}
and the corresponding density matrix is
\begin{equation}\label{MatrixIni}
\rho(k,0) = \frac{\mathbbm{1}+ {\bm{n}} \cdot {\bm{\sigma }}}{2},~~~~~~\bm{n} = \frac{{\left( { - {\delta _1}\cos k  - {\delta _2},0,\omega \sin k } \right)}}{{\sqrt {{{(\omega \sin k )}^2} + {{({\delta _1}\cos k  + {\delta _2})}^2}} }}.
\end{equation}
In the singular case where $\delta_1=\delta_2$ and $k=\pi$, the eigenstates in Eq.~(\ref{eigen1}) is redefined as
\begin{equation}
\begin{aligned}
&{\left. {|+(k,\tau)\rangle} \right|_{{\delta _1} = {\delta _2}, k = \pi}}
= \mathop {\lim }\limits_{k \to \pi^- } \Bigg\{ \mathop {\lim }\limits_{\delta _1 \to \delta _2 }\frac{1}{\sqrt{2}}\begin{bmatrix}\sqrt{1+\frac{\delta}{\Delta}}\\
{\rm sgn}(\frac{\nu}{\Delta})e^{i\phi}\sqrt{1-\frac{\delta}{\Delta}}
\end{bmatrix}\Bigg\} = \frac{1}{\sqrt{2}}\left( {\begin{array}{*{20}{c}}
   1  \\
   {{e^{i\phi }}}  \\
\end{array}} \right),\\
&{\left. {|-(k,\tau)\rangle} \right|_{{\delta _1} = {\delta _2}, k = \pi}}
=\mathop {\lim }\limits_{k \to \pi^- } \Bigg\{ \mathop {\lim }\limits_{\delta _1 \to \delta _2 }\frac{1}{\sqrt{2}}\begin{bmatrix}{\rm sgn}(\frac{\nu}{\Delta})\sqrt{1-\frac{\delta}{\Delta}}\\
-e^{i\phi}\sqrt{1+\frac{\delta}{\Delta}}
\end{bmatrix}\Bigg\} = \frac{1}{\sqrt{2}}\left( {\begin{array}{*{20}{c}}
   1  \\
   {{-e^{i\phi }}}  \\
\end{array}} \right).
\end{aligned}
\label{eigensing}
\end{equation}
Likewise, the initial state in Eq.~(\ref{eq:StateIni}) is redefined as
\begin{equation}
{\left. {|\Psi(k,0)\rangle} \right|_{{\delta _1} = {\delta _2}, k = \pi}}
=\mathop {\lim }\limits_{k \to \pi^- } \Bigg\{ \mathop {\lim }\limits_{\delta _1 \to \delta _2 }\frac{1}{2}\begin{bmatrix}\sqrt{1+\frac{\delta}{\Delta}}+{\rm sgn}(\frac{\nu}{\Delta})\sqrt{1-\frac{\delta}{\Delta}}\\
{\rm sgn}(\frac{\nu}{\Delta})\sqrt{1-\frac{\delta}{\Delta}}-\sqrt{1+\frac{\delta}{\Delta}}
\end{bmatrix}\Bigg\} = \left( {\begin{array}{*{20}{c}}
   1  \\
   0  \\
\end{array}} \right),
\label{eq:StateIniSing}
\end{equation}
and the corresponding Bloch vector in Eq.~(\ref{MatrixIni}) is redefined as
\begin{equation}\label{SingularVector}
{\left. {\bm{n}} \right|_{{\delta _1} = {\delta _2}, k = \pi}} = \mathop {\lim }\limits_{k \to \pi^- } \Bigg[ \mathop {\lim }\limits_{\delta _1 \to \delta _2 }  {\frac{{\left( { - {\delta _1}\cos k - {\delta _2},0,\omega \sin k} \right)}}{{\sqrt {{{(\omega \sin k)}^2} + {{({\delta _1}\cos k + {\delta _2})}^2}} }}} \Bigg]   = (0,0,1).
\end{equation}

Plugging the initial state in Eq.~(\ref{MatrixIni}) into Eq.~(\ref{eq:Qibc}), with the help of Eqs.~(\ref{eq:Hqubit}) and (\ref{eq:E-nu-del}), we find after some algebra that
\begin{equation}
Q_{{\rm IBC}}=-\frac{1}{2\pi}\int_{0}^{\pi}dk\frac{(\omega^{2}-\delta_{1}^{2})\cos(k)-\delta_{1}\delta_{2}}{\left\{ \omega^{2}\sin^{2}(k)+\left[\delta_{1}\cos(k)+\delta_{2}\right]^{2}\right\} ^{3/2}}\omega\sin^{2}(k)\dot{\phi}|_{\tau=0}.
\label{eq:Qibc-2-level-generalramp}
\end{equation}
This expression for $Q_{{\rm IBC}}$, now involving only the $k$ integral from $0$ to $\pi$,
motivates us to restrict $k$ to the regime of $[0,\pi]$ in our actual experiment.

Around the band touching point ($\delta_1=\delta_2, k=\pi$), the numerator of the integrand in Eq.~(\ref{eq:Qibc-2-level-generalramp}) approaches zero as $|k-\pi|^2$, whereas the denominator of the same integrand approaches zero as $|k-\pi|^3$. Therefore, the integrand itself approaches zero as $|k-\pi|^{-1}$ and its integration over $k$ hence yields
a logarithmic divergence around $k=\pi$.  This explains the origin of the divergence in the theoretical pumping curve shown in
Fig.~\ref{final}(a) of the main text.

For the specific linear ramp case, $\phi=2\pi\tau$, the above equation becomes
\begin{equation}
Q_{{\rm IBC}}=-\int_{0}^{\pi}dk\frac{(\omega^{2}-\delta_{1}^{2})\cos(k)-\delta_{1}\delta_{2}}{\left\{ \omega^{2}\sin^{2}(k)+\left[\delta_{1}\cos(k)+\delta_{2}\right]^{2}\right\} ^{3/2}}\omega\sin^{2}(k).
\label{eq:Qibc-2-level}
\end{equation}
In Fig.~\ref{final}(a) of the main text, the gray theoretical curve (for $T\rightarrow\infty$) is obtained by evaluating this term with $\omega=2\delta_{1}$ and $\delta_{2}\in[0,2\delta_{1}]$.

Additionally, since both $\rho_{-+}(k,\tau)$ and $v_{+-}(k,\tau)$ are real at $\tau=0$, the term $Q_{\rm{NG}}$ vanishes according to Eq.~(\ref{eq:Qts}). To summarize, for our
initial state $|\Psi(k,0)\rangle$, the charge pumping over an adiabatic
cycle is solely given by $Q_{{\rm IBC}}$, the contribution due to
interband coherence in the initial state.

\subsection{1.3 Reflection symmetry of the velocity expectation value}
For the specific initial state in Eq.~(\ref{MatrixIni}), the velocity expectation value has reflection symmetry in $k$, i.e., $\langle v(k,\tau) \rangle = \langle v(-k,\tau) \rangle$. The proof is as follows.

We turn to the rotating frame that rotates around the $z$ axis according to $\phi(\tau)$, or in other words, we apply the rotating transformation characterized by the rotation operator $R=e^{-i\phi(\tau)\sigma_z/2}$. The Hamiltonian, the evolution operator, the density operator, and the velocity operator in this rotating frame are
\begin{equation}\label{rotate}
\begin{aligned}
 &\widetilde{H}(k,\tau) = {R^\dag }H(k,\tau)R - \frac{i}{T}{R^\dag }\frac{d}{{d\tau}}R = \nu \frac{{{\sigma _x}}}{2} + \left( {\delta - \frac{1}{T}\frac{{d\phi }}{{d\tau}}} \right)\frac{{{\sigma _z}}}{2},\\
 &\widetilde{U}(k,\tau)=\mathcal{T} {{\exp \left[ - i\int_0^\tau\widetilde{H}(k,\tau)d\tau'\right]}}, \\
 &\widetilde{\rho}(k,\tau) = {R^\dag }\rho(k,\tau) R = \widetilde{U}(k,\tau)\widetilde\rho(k,0)\widetilde{U}(k,\tau)^\dag,\\
 &\widetilde{v}(k,\tau) = {R^\dag }v(k,\tau)R = \omega \cos k \frac{{{\sigma _x}}}{2} - {\delta _1}\sin k \frac{{{\sigma _z}}}{2}, \\
\end{aligned}
\end{equation}
where $\nu\equiv\omega\sin k$ and $\delta\equiv\delta_{1}\cos k +\delta_{2}$ are the same as Eq.~(\ref{eq:E-nu-del}).
One can observe that
$$
\begin{aligned}
&{\sigma _z}\widetilde{H}(k,\tau){\sigma _z} = \widetilde{H}(-k,\tau),~~~~~~{\sigma _z}\widetilde{U}(k,\tau){\sigma _z} = \widetilde{U}(-k,\tau),\\
&{\sigma _x}\widetilde{v}(k,\tau){\sigma _x} = \widetilde{v}(-k,\tau),~~~~~~{\sigma_x}\widetilde{\rho}(k,0){\sigma _x} = \widetilde{\rho}(-k,0),\\
&{\sigma _y}\widetilde{\rho}(k,0){\sigma _y} =  - \widetilde{\rho}(k,0),~~~~~~{\sigma _y}\widetilde{v}(k,\tau){\sigma _y} =  - \widetilde{v}(k,\tau).\\
\end{aligned}
$$
The above relations entail
$$
\begin{aligned}
\left\langle \widetilde{v}(-k,\tau) \right\rangle  &= {\rm{tr}}[\widetilde{\rho}(-k,\tau)\widetilde{v}(-k,\tau)] = {\rm{tr}}\left[ \widetilde{U}(-k,\tau)\widetilde{\rho}(-k,0)\widetilde{U}(-k,\tau)\widetilde{v}(-k,\tau) \right] \\
&= {\rm{tr}}\left[ {{\sigma _z}\widetilde{U}(k,\tau){\sigma _z}{\sigma _x}\widetilde{\rho}(k,0){\sigma _x}{\sigma _z}\widetilde{U}(k,\tau)^\dag{\sigma _z}{\sigma _x}\widetilde{v}(k,\tau){\sigma _x}} \right] \\
&= {\rm{tr}}\left[ \widetilde{U}(k,\tau){\sigma _y}\widetilde{\rho}(k,0){\sigma _y}\widetilde{U}(k,\tau)^\dag{\sigma _y}\widetilde{v}(k,\tau){\sigma _y} \right]\\
&= {\rm{tr}}\left[ \widetilde{U}(k,\tau)\widetilde{\rho}(k,0)\widetilde{U}(k,\tau)^\dag \widetilde{v}(k,\tau) \right] = {\rm{tr}}[\widetilde{\rho}(k,\tau)\widetilde{v}(k,\tau)] = \left\langle \widetilde{v}(k,\tau) \right\rangle.\\
\end{aligned}
$$
By taking ${\rm{tr}}[\widetilde{\rho}(k,\tau)\widetilde{v}(k,\tau)]={\rm{tr}}[\rho(k,\tau)v(k,\tau)]$ into account, one finally obtains $\langle v(k,\tau) \rangle = \langle v(-k,\tau) \rangle$. Therefore, for the specific initial state in Eq.~(\ref{MatrixIni}), only half of the first Brillouin zone is enough for evaluating the pumped charge $Q$ in Eq.~(\ref{eq:Q}).

\subsection{1.4 Detailed solution for the linear ramp case}
For the linear driving protocol $\phi=2\pi\tau$, the Hamiltonian in the rotating frame is
\begin{equation}\label{linearrotate}
 \widetilde{H}(k) = \nu \frac{{{\sigma _x}}}{2} + \left( {\delta - \frac{2\pi}{T}} \right)\frac{{{\sigma _z}}}{2},
\end{equation}
according to Eq.~(\ref{rotate}). Note that this Hamiltonian is time-independent. The eigenenergies and eigenstates of $\widetilde{H}(k)$ are
\begin{equation}
\widetilde{E}_{\pm}(k)=\pm\frac{1}{2}\sqrt{\nu^{2}+\widetilde{\delta}^{2}},\qquad
|\widetilde{+}(k)\rangle=\frac{1}{\sqrt{2}}\begin{bmatrix}\sqrt{1+\frac{\widetilde{\delta}}{\widetilde{\Delta}}}\\
{\rm sgn}(\nu)\sqrt{1-\frac{\widetilde{\delta}}{\widetilde{\Delta}}}
\end{bmatrix},\qquad|\widetilde{-}(k)\rangle=\frac{1}{\sqrt{2}}\begin{bmatrix}{\rm sgn}(\nu)\sqrt{1-\frac{\widetilde{\delta}}{\widetilde{\Delta}}}\\
-\sqrt{1+\frac{\widetilde{\delta}}{\widetilde{\Delta}}}
\end{bmatrix},
\end{equation}
with
\begin{equation}
\widetilde\delta\equiv\delta-\frac{2\pi}{T},\qquad\widetilde\Delta\equiv \widetilde{E}_{+}(k)-\widetilde{E}_{-}(k)=\sqrt{\nu^{2}+\widetilde\delta^{2}}.\label{eq:rotdel}
\end{equation}

The shift of peaks in the experimental observation of charge pumping
may be investigated from the spectrum of $\widetilde{H}(k)$, which is
gapless at $k=\pi$ if the two frequencies $\delta_{1}$ and $\delta_{2}$
satisfy the relation $\delta_{2}=\delta_{1}+2\pi/T$. So for a finite
$T$, the position of peak will shift from $\delta_{2}/\delta_{1}=1$
to $\delta_{2}/\delta_{1}=1+\frac{2\pi}{\delta_{1}T}>1$. With the
increasing of $T$, the peak will shift gradually to left, until being
coincide with the adiabatic result at $\delta_{2}=\delta_{1}$ when
$T\rightarrow\infty$. In the following we give more detailed calculations
to support this argument.

Thanks to the time-independence of the Hamiltonian in Eq.~(\ref{linearrotate}), the evolution of a state in this rotating frame can be solved analytically.
Due to ${\rm{tr}}[\widetilde{\rho}(k,\tau)\widetilde{v}(k,\tau)]={\rm{tr}}[\rho(k,\tau)v(k,\tau)]$, we can evaluate Eq.~(\ref{eq:Q}) in the rotating frame. After some algebra we find
\begin{alignat}{1}
Q & =Q_{\rm{st}}+Q_{\rm{os}},\\
Q_{\rm{st}} & \equiv\frac{1}{4\pi}\int_{-\pi}^{\pi}dkT\frac{\partial\tilde{\Delta}}{\partial k}\left[|\langle\widetilde{+}|\Psi(k,0)\rangle|^{2}-|\langle\widetilde{-}|\Psi(k,0)\rangle|^{2}\right]\label{eq:Qbardef},\\
Q_{\rm{os}} & \equiv\frac{1}{\pi}\int_{-\pi}^{\pi}dk\frac{\langle\Psi(k,0)|\widetilde{+}\rangle\langle\widetilde{+}|\frac{\partial\tilde{H}}{\partial k}|\widetilde{-}\rangle\langle\widetilde{-}|\Psi(k,0)\rangle}{\tilde{\Delta}}\sin(\tilde{\Delta}T),\label{eq:Qtiddef}
\end{alignat}
for the initial states given by Eq.~(\ref{eq:StateIni}).
As we will show in the following, $Q_{\rm{st}}$ is the stationary
part of $Q$, and it will converge to $Q_{{\rm IBC}}$ given by Eq.~(\ref{eq:Qibc-2-level}) in the adiabatic limit. On the contrary,
the integrand of $Q_{\rm{os}}$ is an oscillating function of $k$.
When $T$ is large, the integrand of $Q_{\rm{os}}$ will oscillate
fast with respect to $k$, and its contribution to $Q$ after integrating
over $k$ is at least of the order of $1/T$, which will finally vanish in the adiabatic limit.

Straightforward calculations yield
\begin{alignat}{1}
Q_{\rm{st}} & =-\int_{0}^{\pi}dk\frac{\nu(\nu\partial_{k}\nu+\delta\partial_{k}\delta)}{\Delta\tilde{\Delta}^{2}}\nonumber \\
 & =-\int_{0}^{\pi}dk\frac{\left[\left(\omega^{2}-\delta_1^2\right)\cos(k)-\delta_1\delta_{2}\right]\omega\sin^{2}(k)}{\sqrt{\omega^{2}\sin^{2}(k)+\left[\delta_1\cos(k)+\delta_{2}\right]^{2}}\left\{ \omega^{2}\sin^{2}(k)+\left[\delta_1\cos(k)+\delta_{2}-\frac{2\pi}{T}\right]^{2}\right\} }.\label{eq:Qbar}
\end{alignat}
Comparing this with Eq.~(\ref{eq:Qibc-2-level}) for the charge pumping due to interband coherence, we find that in the adiabatic limit ($T\rightarrow\infty$), $Q_{\rm{st}}$ will converge to $Q_{{\rm IBC}}$, i.e.,
\begin{equation}
\lim_{T\rightarrow\infty}Q_{\rm{st}}=Q_{{\rm IBC}}.
\end{equation}
In the large $T$ regime, $Q_{\rm{st}}$ will approach to $Q_{{\rm IBC}}$ algebraically, with leading correction of order $1/T$. Also we
note that the integrand of $Q_{\rm{st}}$ does not diverge at $k=\pi$ for either $\delta_{2}=\delta_{1}$ or $\delta_{2}=\delta_{1}+2\pi/T$.
This implies that in practice the peak is smooth and has a finite height.


Performing similar calculations, the oscillatory part $Q_{\rm{os}}$ in Eq.~(\ref{eq:Qtiddef}) is found to be
\begin{equation}
Q_{\rm{os}}=-\frac{1}{2\pi}\int_{0}^{\pi}dk\frac{\omega\left[\delta_{1}+\left(\delta_{2}-\frac{2\pi}{T}\right)\cos(k)\right]\left(\tilde{\delta}\delta+\nu^{2}\right)}{\Delta\tilde{\Delta}^{3}}\sin(\tilde{\Delta}T)
\end{equation}
The integrand of this integral also does not diverge at $k=\pi$ for either $\delta_{2}=\delta_{1}$ or $\delta_{2}=\delta_{1}+2\pi/T$.
Furthermore, when $T$ is large, $\sin(\tilde{\Delta}T)$ is a fast oscillating function with respect to $k$.
Its integral over $k$ is therefore at least of the order of $1/T$.
So in the adiabatic limit we will have $\lim_{T\rightarrow\infty}Q_{\rm{os}}=0$.
In the large $T$ regime, $Q_{\rm{os}}$ will decrease with the increase of $T$.
Its leading contribution to $Q$ is also of the order of $1/T$.

In Fig.~\ref{final}(a) of the main text, the orange and green curves related to the linear ramp $\phi=2\pi\tau$ are obtained by evaluating Eq.~(\ref{eq:Q}) based on the exact solutions of the Schr\"{o}dinger equation with the initial states given by Eq.~(\ref{eq:StateIni}).
For the blue curve related to the protocol $\phi=2\pi\tau^{2}$, the corresponding rotating-frame Hamiltonian is also time dependent. So this curve is found by first solving the Schr\"{o}dinger equation numerically with the initial state~(\ref{eq:StateIni}) and then computing the charge pumping with Eq.~(\ref{eq:Q}).


\section{2. Experiment}

\subsection{2.1 Experimental setup}
The experiment is performed on an NV center in a \{100\}-face bulk diamond synthesized by chemical vapor deposition (CVD).
The nitrogen impurity is less than 5 ppb and the abundance of $^{13}$C is at the natural level of about 1.1\%. The dephasing time of the NV electron spin is 1.7 $\mu$s.
The NV center is optically addressed by a home-built confocal microscope. Green laser is used for optical excitation.
The laser beam is released and cut off by an acousto-optic modulator (power leakage ratio $\sim1/1000$). To reduce the laser leakage further, the beam passes twice through the acousto-optic modulator. The laser is focused into the diamond by an oil objective (60*O, NA 1.42). The phonon sideband fluorescence with the wavelength between 650 and 800 nm is collected by the same oil objective and finally detected by an avalanche photodiode with a counter card. A solid immersion lens etched on the diamond by focused ion beam enhances the fluorescence counting rate up to 400 thousand counts per second.
The microwaves generated by an arbitrary waveform generator (AWG) pass a 6 dB attenuator and then strengthened by a power amplifier. Finally, the microwaves are radiated to the NV center from a coplanar waveguide. The magnetic field is supplied by a permanent magnet mounted on a manual translation stage.

\subsection{2.2 Calibration}
The magnitude of the transverse field is $\omega\sin k$ during the evolution period. In order to feed the microwaves with proper amplitude to the NV center, we calibrate the Rabi frequency as a function of the AWG's output amplitude. The calibration is done by performing conventional Rabi oscillation experiments with various output amplitudes. We fit the experimental data of Rabi oscillation associated with each output amplitude $V$ to extract the corresponding Rabi frequency $\omega_{\rm R}$, and then fit the Rabi frequency using $\omega_{\rm R}=a V^b$ with $a$ and $b$ being the coefficients to be determined. The relation between the Rabi frequency and the AWG's output amplitude is thus obtained. Such calibration is carried out hourly to guard against the drift of experimental conditions.

\subsection{2.3 Pulse sequence}
After the qubit is polarized to the state $\left| {{\psi _0}} \right\rangle={(1,0)^{\rm{T}}}$ by a green laser pulse, a resonant microwave pulse is applied to prepare the initial state.
In the laboratory frame, the Hamiltonian of the qubit irradiated by the pulse is
\begin{equation}\label{PrepLab}
H_{{\rm{ini}}}^{{\rm{lab}}} = \frac{\omega _0}{2}\sigma _z + \omega_1 \cos ( {\omega _0} t +\varphi_{\rm{ini}})\sigma _x.
\end{equation}
where the first term on the right-hand side is the static component of the Hamiltonian with $\omega _0$ being the resonant frequency,
and the second term accounts for the effect of the microwaves with $\omega_1$, $\varphi_{\rm{ini}}$, and $t$ being the Rabi frequency, the initial phase, and the time starting from zero, respectively.
The value of $\varphi_{\rm{ini}}$ depends on ${\delta _1}\cos k  + {\delta _2}$ as $\varphi_{\rm{ini}}=-{\pi}/{2}$ for ${\delta _1}\cos k  + {\delta _2}\ge0$
and $\varphi_{\rm{ini}}={\pi}/{2}$ for ${\delta _1}\cos k  + {\delta _2}<0$.
In the rotating frame that rotates around the $z$ axis with the angular frequency $\omega_0$, or to put it another way, under the rotating transformation characterized by the rotation operator $R_{\rm{ini}} = {e^{ - i{\omega _0} t{\sigma _z}/2}}$,
the Hamiltonian in Eq.~(\ref{PrepLab}) is transformed to
\begin{equation}\label{PrepRot}
H_{{\rm{ini}}}^{{\rm{rot}}} = {R_{\rm{ini}}^\dag }H_{{\rm{ini}}}^{{\rm{lab}}}R_{\rm{ini}} - i{R_{\rm{ini}}^\dag }\frac{d}{{dt}}R_{\rm{ini}} = \frac{\omega _1}{2} ( \sigma _x\cos \varphi_{\rm{ini}} +\sigma _y\sin \varphi_{\rm{ini}}),
\end{equation}
where the second equality is based on the rotating wave approximation.
The pulse lasts for $t_{\rm{ini}}=\alpha/\omega_1$, where $\alpha$ is the inclination angle of the initial state. From Eq.~(\ref{MatrixIni}) one can see that, in usual cases,
\begin{equation}\label{IniPolarAngle}
\alpha=\arccos\frac{ \omega \sin k }{{\sqrt {{{(\omega \sin k )}^2} + {{({\delta _1}\cos k  + {\delta _2})}^2}} }}.
\end{equation}
From Eq.~(\ref{SingularVector}) one can see that, in the singular case where $\delta_1=\delta_2$ and $k=\pi$, the angle $\alpha$ equals zero.
Therefore, after the pulse, the state of the qubit in the rotating frame is
\begin{equation}\label{IniStateRot}
\left| {\psi _{{\rm{ini}}}^{{\rm{rot}}}} \right\rangle = e^{-iH_{{\rm{ini}}}^{{\rm{rot}}}t_{\rm{ini}}} \left| {{\psi _0}} \right\rangle = \left| {\Psi {\rm{(}}k,0{\rm{)}}} \right\rangle,
\end{equation}
which is our desired initial state. In the laboratory frame, this state immediately after the pulse is expressed as
\begin{equation}\label{IniStateLab}
\left| {\psi _{{\rm{ini}}}^{{\rm{lab}}}} \right\rangle  = {e^{ - i{\omega _0} t_{\rm{ini}} {\sigma _z}/2}} \left| {\psi _{{\rm{ini}}}^{{\rm{rot}}}} \right\rangle.
\end{equation}

Next, a microwave pulse is applied to build the model Hamiltonian in Eq.~(\ref{eq:Hqubit}). In most cases, this pulse is off-resonant.
In the laboratory frame, the Hamiltonian of the qubit irradiated by the pulse is
\begin{equation}\label{PumpLab}
H_{{\rm{pump}}}^{{\rm{lab}}} = \frac{\omega _0}{2}\sigma _z + \omega \sin k \cos \left[ {({\omega _0} - {\delta _1}\cos k  - {\delta _2})t + \phi \left(\frac{t}{T}\right)} + \omega _0 t_{\rm{ini}}\right]\sigma _x,
\end{equation}
where $t$ is the time starting from zero.
In the rotating frame with the rotation operator $R_{\rm{pump}} = {e^{ - i[({\omega _0} - {\delta _1}\cos k  - {\delta _2})t + \omega _0 t_{\rm{ini}}]{\sigma _z}/2}}$, the Hamiltonian in Eq.~(\ref{PumpLab}) is transformed to our target Hamiltonian, namely,
\begin{equation}\label{PumpRot}
H_{{\rm{pump}}}^{{\rm{rot}}} = {R_{\rm{pump}}^\dag }H_{{\rm{pump}}}^{{\rm{lab}}}R_{\rm{pump}} - i{R_{\rm{pump}}^\dag }\frac{d}{{dt}}R_{\rm{pump}} = \frac{\omega \sin k}{2} \left[ {\cos \phi \left(\frac{t}{T}\right){{\sigma _x}} + \sin \phi \left(\frac{t}{T}\right){{\sigma _y}}} \right]+\frac{{\delta _1}\cos k+{\delta _2}}{2}\sigma _z,
\end{equation}
where the second equality is based on the rotating wave approximation.
In this rotating frame, the state in Eq.~(\ref{IniStateLab}) is rewritten as
\begin{equation}\label{IniStateRot}
R_{\rm{pump}}^\dag(t=0)\left| {\psi _{{\rm{ini}}}^{{\rm{lab}}}} \right\rangle=\left| {\psi _{{\rm{ini}}}^{{\rm{rot}}}} \right\rangle  = \left| {\Psi {\rm{(}}k,0{\rm{)}}} \right\rangle,
\end{equation}
which has the same form as in the rotating frame defined by $R_{\rm{ini}}$.
The pulse lasts for some duration duration $t_{\rm e}\in [0,T]$, which is a sampling point in time.
Assume that the state of the qubit immediately after the pulse is $\left| {\psi _{{\rm{pump}}}^{{\rm{rot}}}} \right\rangle$ in this rotating frame.
In the laboratory frame, the state is expressed as
\begin{equation}\label{PumpStateLab}
\left| {\psi _{{\rm{pump}}}^{{\rm{lab}}}} \right\rangle={e^{ - i[({\omega _0} - {\delta _1}\cos k  - {\delta _2})t_{\rm e} + \omega _0 t_{\rm{ini}}]{\sigma _z}/2}}\left| {\psi _{{\rm{pump}}}^{{\rm{rot}}}} \right\rangle.
\end{equation}

Finally, a resonant microwave pulse is applied to assist measurement.
In the laboratory frame, the Hamiltonian of the qubit irradiated by the pulse is
\begin{equation}\label{FinLab}
H_{{\rm{fin}}}^{{\rm{lab}}} = \frac{\omega _0}{2}\sigma _z + \omega_1 \cos \left[ {{\omega _0}t + ({\omega _0} - {\delta _1}\cos k  - {\delta _2})t_{\rm e} + \phi \left(\frac{t_{\rm e}}{T}\right)} + \omega _0 t_{\rm{ini}} + \varphi_{\rm{fin}}\right]\sigma _x,
\end{equation}
where $\varphi_{\rm{fin}}=-\pi/2$ for $\cos k \ge 0$ and $\varphi_{\rm{fin}}=\pi/2$ for $\cos k<0$, and $t$ is also the time starting from zero.
In the rotating frame with the rotation operator $R_{\rm{fin}} = {e^{ - i[{\omega _0} t+({\omega _0} - {\delta _1}\cos k  - {\delta _2})t_{\rm e}+\omega _0 t_{\rm{ini}}]{\sigma _z}/2}}$,
the Hamiltonian in Eq.~(\ref{FinLab}) is transformed to
\begin{equation}\label{ReadRot}
H_{{\rm{fin}}}^{{\rm{rot}}} = {R_{\rm{fin}}^\dag }H_{{\rm{fin}}}^{{\rm{lab}}}R_{\rm{fin}} - i{R_{\rm{fin}}^\dag }\frac{d}{{dt}}R_{\rm{fin}}
= \frac{\omega _1}{2} \left\{ \sigma _x\cos \left[\phi \left(\frac{t_{\rm e}}{T}\right)+\varphi_{\rm{fin}}\right] +\sigma _y\sin\left[ \phi \left(\frac{t_{\rm e}}{T}\right)+\varphi_{\rm{fin}}\right]\right\},
\end{equation}
where the second equality is based on the rotating wave approximation.
In this rotating frame, the state in Eq.~(\ref{PumpStateLab}) is rewritten as
\begin{equation}\label{PumpStateRot}
R_{\rm{fin}}^\dagger(t=0)\left| {\psi _{{\rm{pump}}}^{{\rm{lab}}}} \right\rangle = \left| {\psi _{{\rm{pump}}}^{{\rm{rot}}}} \right\rangle,
\end{equation}
which has the same form as in the rotating frame defined by $R_{\rm{pump}}$.
The pulse lasts for $t_{\rm{ini}}=\beta/\omega_1$, where
\begin{equation}\label{VelPolarAngle}
\beta=\arccos\frac{-\delta_1 \sin k }{{\sqrt {{{(\omega \cos k )}^2} + {{({\delta_1}\sin k})}^2}} }.
\end{equation}
is the inclination angle of the direction of the velocity operator $v=\partial_k H=\omega \cos k (\cos\phi~\sigma_x+\sin\phi~\sigma_y)/2 - \delta_1\sin k~\sigma_z/2$.
After the pulse, laser illumination is carried out to realize the measurement of $\sigma_z$.
The combined effect of the final microwave pulse and its subsequent laser illumination amounts to the measurement of
\begin{equation}\label{VelMeas}
e^{iH_{{\rm{fin}}}^{{\rm{rot}}}t_{\rm{fin}}} \sigma_z e^{-iH_{{\rm{fin}}}^{{\rm{rot}}}t_{\rm{fin}}}
= \frac{v}{\left\|v\right\|}.
\end{equation}


\subsection{2.4 Experimental data analysis}
As shown in Fig.~\ref{structure}(c) of the main text, the spin state is read out during the latter laser pulse and there are two counting windows. Such sequence is iterated at least a hundred thousand times. The total photon count recorded by the first (second) window during these iterations is regarded as signal (reference) and denoted by $s$ ($r$). The raw experimental data is $x=s/r$. To normalize the data, a conventional Rabi oscillation is performed alongside. We fit the raw data of the Rabi oscillation using the function $x=x_0+a\cos(\omega_{\rm R} t+\varphi)$, and then normalize the experimental data as $x_{\rm{n}}=(x-x_0)/a$. The data thus normalized represent the expectation value of $\sigma_z$. In the experiment, the sampling interval in time $t=T\tau$ is 10 ns. For $\delta_2/\delta_1=$0, 0.5, 1.5, and 2, the sampling interval in $k$ is $\pi/18$. For other values of $\delta_2/\delta_1$ around the phase transition point $\delta_2/\delta_1=1$, the sampling interval in $k$ is $\pi/18$ when $0\le k\le8\pi/9$ and is $\pi/90$ when $8\pi/9\le k\le\pi$. The numerical integration is based on Simpson's rule.

\subsection{2.5 Experimental data}
The complete data that support the final results in Fig.~\ref{final}(a) of the main text are as follows.
~\\~\\~\\~\\~\\~\\~\\~\\~\\~\\~\\~\\~\\~\\~\\~\\~\\~\\~\\~\\~\\~\\~\\~\\~\\~\\~\\~\\~\\~\\~\\~\\~\\~\\~\\~\\~\\~\\~\\
\begin{figure}
\includegraphics[width=1\columnwidth]{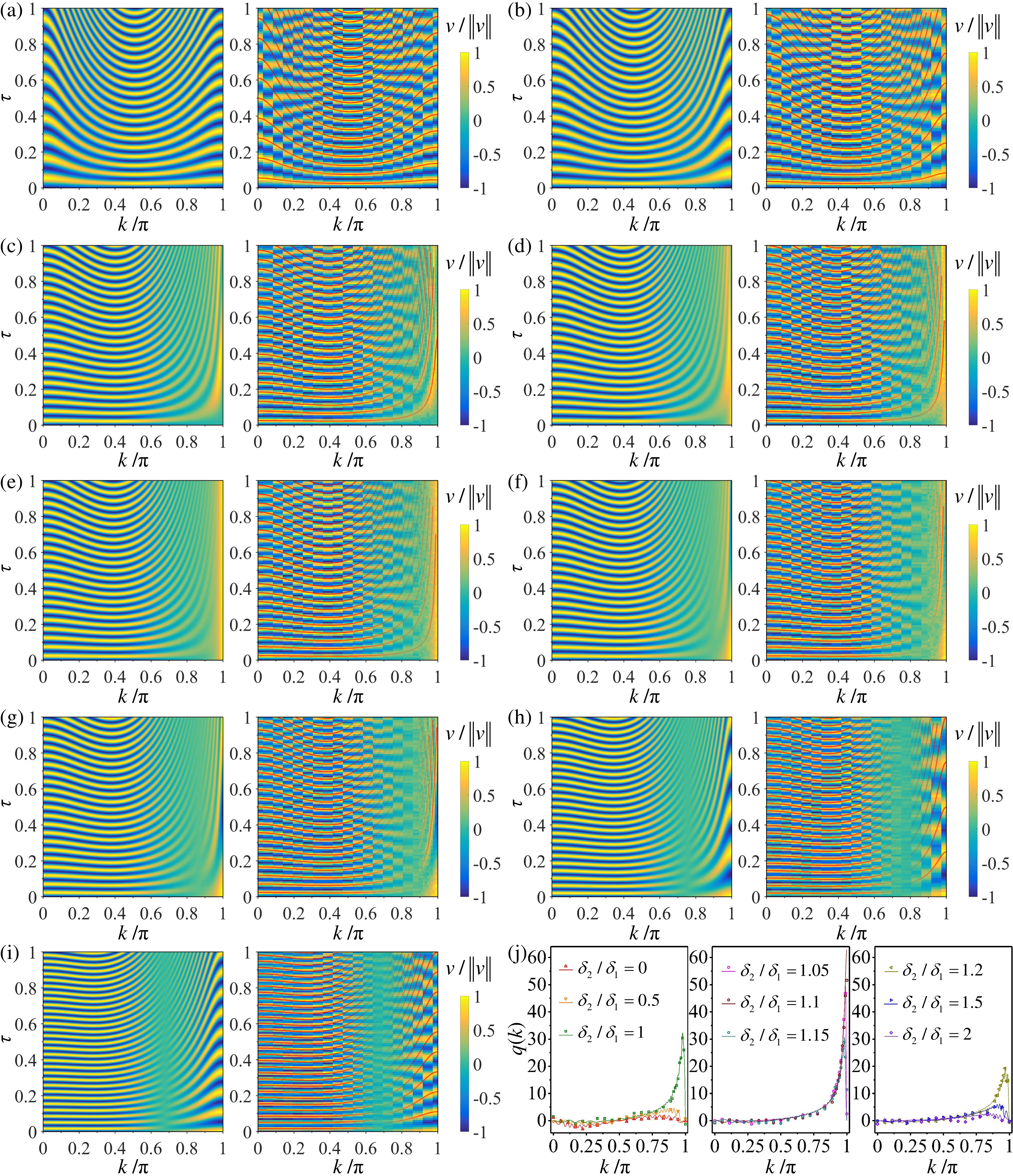}
\caption{Normalized velocity expectation values and pumped charge per each $k$ for $\phi(\tau)=2 \pi \tau$ with $T=1$ $\rm\mu$s.
(a-i) Normalized velocity expectation values $\langle v \rangle / \left\|v\right\|$ as a function of the synthetic quasimomentum $k$ and the scaled time $\tau$ for $\delta_2/\delta_1=$0, 0.5, 1, 1.05, 1.1, 1.15, 1.2, 1.5, and 2, respectively. The calculations based on the Schr\"{o}dinger equation are on the left and the experimental data are on the right. The red curves in experimental contour maps are guides to the eye to clarify the patterns in the color map. These guidelines are the crest lines in the patterns of the calculated $\langle v \rangle / \left\|v\right\|$. The transparency of the guidelines is related to the values of $\langle v \rangle / \left\|v\right\|$ and reflects the amplitude of oscillation.
(j) Pumped charge $q(k)$ contributed from each quasimomentum $k$ for $\delta_2/\delta_1=$0, 0.5, 1, 1.05, 1.1, 1.15, 1.2, 1.5, and 2.
Symbols represent the experimental data and curves represent the calculation.
   }
    \label{linearone}
\end{figure}

\begin{figure}
\includegraphics[width=1\columnwidth]{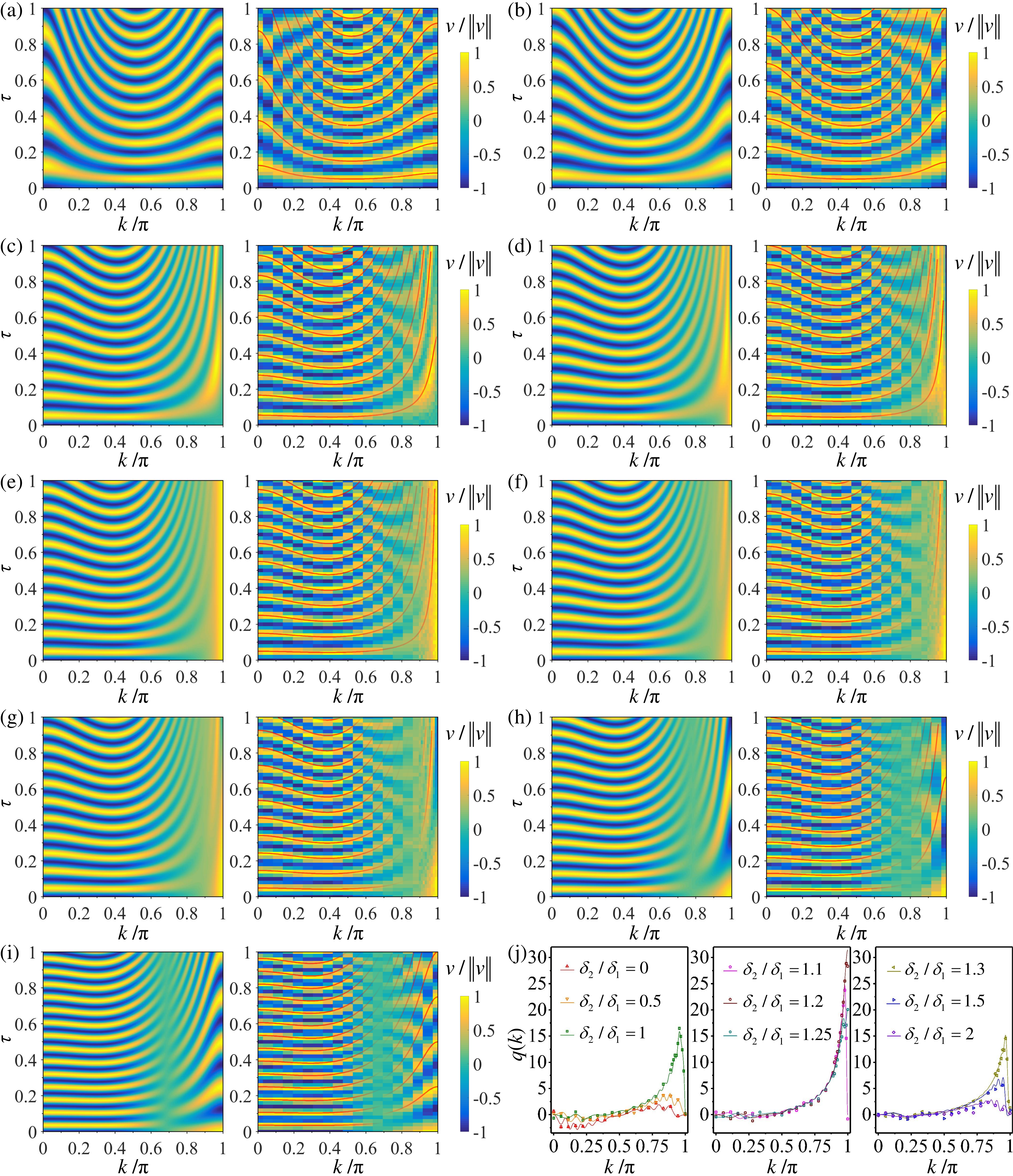}
\caption{Normalized velocity expectation values and pumped charge per each $k$ for $\phi(\tau)=2 \pi \tau$ with $T=0.5$ $\rm\mu$s.
(a-i) Normalized velocity expectation values $\langle v \rangle / \left\|v\right\|$ as a function of the synthetic quasimomentum $k$ and the scaled time $\tau$ for $\delta_2/\delta_1=$0, 0.5, 1, 1.1, 1.2, 1.25, 1.3, 1.5, and 2, respectively. The calculations based on the Schr\"{o}dinger equation are on the left and the experimental data are on the right. The red curves in experimental contour maps are guides to the eye to clarify the patterns in the color map. These guidelines are the crest lines in the patterns of the calculated $\langle v \rangle / \left\|v\right\|$. The transparency of the guidelines is related to the values of $\langle v \rangle / \left\|v\right\|$ and reflects the amplitude of oscillation.
(j) Pumped charge $q(k)$ contributed from each quasimomentum $k$ for $\delta_2/\delta_1=$0, 0.5, 1, 1.1, 1.2, 1.25, 1.3, 1.5, and 2.
Symbols represent the experimental data and curves represent the calculation.
   }
    \label{linearhalf}
\end{figure}

\begin{figure}
\includegraphics[width=1\columnwidth]{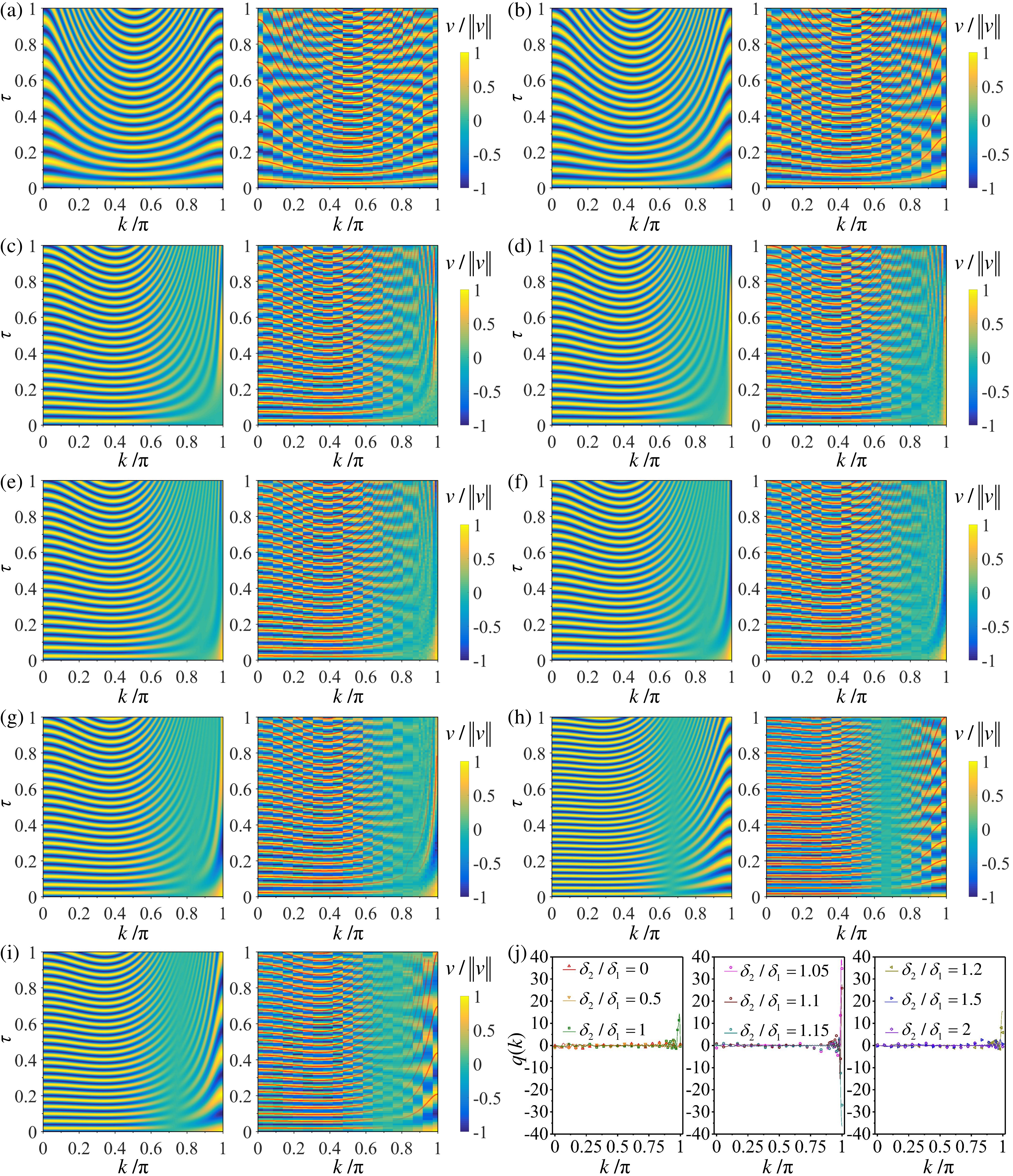}
\caption{Normalized velocity expectation values and pumped charge per each $k$ for $\phi(\tau)=2 \pi \tau^2$ with $T=1$ $\rm\mu$s.
(a-i) Normalized velocity expectation values $\langle v \rangle / \left\|v\right\|$ as a function of the synthetic quasimomentum $k$ and the scaled time $\tau$ for $\delta_2/\delta_1=$0, 0.5, 1, 1.05, 1.1, 1.15, 1.2, 1.5, and 2, respectively. The calculations based on the Schr\"{o}dinger equation are on the left and the experimental data are on the right. The red curves in experimental contour maps are guides to the eye to clarify the patterns in the color map. These guidelines are the crest lines in the patterns of the calculated $\langle v \rangle / \left\|v\right\|$. The transparency of the guidelines is related to the values of $\langle v \rangle / \left\|v\right\|$ and reflects the amplitude of oscillation.
(j) Pumped charge $q(k)$ contributed from each quasimomentum $k$ for $\delta_2/\delta_1=$0, 0.5, 1, 1.05, 1.1, 1.15, 1.2, 1.5, and 2.
Symbols represent the experimental data and curves represent the calculation.
   }
    \label{quadratic}
\end{figure}


\begin{thebibliography}{99}
\bibitem{NV}
The Hamiltonian of the electronic ground state of the NV center with a static magnetic field $B$ applied along the NV axis (also the $z$ axis) is $H_{\rm{NV}} = D S_z^2+\gamma B S_z$,
where $S_z$ is the angular momentum operator for spin-1, $D=2\pi\times2870$~MHz is the zero-field splitting, and $\gamma=2\pi\times2.8$~MHz/G is the gyromagnetic ratio of the NV electron.
\end{thebibliography}
\end{document}